\documentclass[fleqn,10pt,twocolumn]{wlscirep}
\usepackage[utf8]{inputenc}
\usepackage{bm,latexsym,mathrsfs,enumerate,color}
\usepackage[mathcal]{euscript}
\usepackage{upgreek}
\usepackage{mathtools}
\usepackage{pgf}
\usepackage{pgfplots}
\usepackage{graphicx}
\usepackage{xifthen}

\usepackage{media9}
\addmediapath{./movies}

\usepackage[textwidth=15mm]{todonotes}
\newcounter{todocounter}

\renewcommand{\vec}[1]{\bm{#1}}
\newenvironment{methods}{
\scriptsize %
\setcounter{equation}{0}
}{}

\usepackage[normalem]{ulem}

\title{Rashba Torque Driven Domain Wall Motion in Magnetic Helices\\
{\normalsize November 25, 2015}
}

\author[1,*]{Oleksandr V. Pylypovskyi}
\author[1,$\dagger$]{Denis D. Sheka}
\author[2,$\ddagger$]{Volodymyr P. Kravchuk}
\author[2,3,$\S$]{Kostiantyn~V.~Yershov}
\author[4,5,$\P$]{Denys Makarov}
\author[2,**]{Yuri Gaididei}
\affil[1]{Taras Shevchenko National University of Kyiv, 01601 Kyiv, Ukraine}
\affil[2]{Bogolyubov Institute for Theoretical Physics of the National Academy of Sciences of Ukraine, 03680 Kyiv, Ukraine}
\affil[3]{National University of ``Kyiv-Mohyla Academy'', 04655 Kyiv, Ukraine}
\affil[4]{Institute of Ion Beam Physics and Materials Research, Helmholtz-Zentrum Dresden-Rossendorf e. V., 01328 Dresden, Germany}
\affil[5]{Institute for Integrative Nanosciences, IFW Dresden, 01069 Dresden, Germany}
\affil[*]{engraver@univ.net.ua}
\affil[$\dagger$]{sheka@univ.net.ua}
\affil[$\ddagger$]{vkravchuk@bitp.kiev.ua}
\affil[$\S$]{yershov@bitp.kiev.ua}
\affil[$\P$]{d.makarov@hzdr.de}
\affil[**]{ybg@bitp.kiev.ua}


\begin{abstract}
Manipulation of the domain wall propagation in magnetic wires is a key practical task for a number of devices including racetrack memory and magnetic logic. Recently, curvilinear effects emerged as an efficient mean to impact substantially the statics and dynamics of magnetic textures. Here, we demonstrate that the curvilinear form of the exchange interaction of a magnetic helix results in an effective anisotropy term and Dzyaloshinskii--Moriya interaction with a complete set of Lifshitz invariants for a one-dimensional system. In contrast to their planar counterparts, the geometrically induced modifications of the static magnetic texture of the domain walls in magnetic helices offer unconventional means to control the wall dynamics relying on spin-orbit Rashba torque. The chiral symmetry breaking due to the Dzyaloshinskii--Moriya interaction leads to the opposite directions of the domain wall motion in left- or right-handed helices. Furthermore, for the magnetic helices, the emergent effective anisotropy term and Dzyaloshinskii--Moriya interaction can be attributed to the clear geometrical parameters like curvature and torsion offering intuitive understanding of the complex curvilinear effects in magnetism.
\end{abstract}

\begin{document}
{  
\onecolumn

\flushbottom
\maketitle

\thispagestyle{empty}
Assessing spin textures of three-dimensionally curved magnetic thin films\cite{Albrecht05, Ulbrich06, Hertel13a}, hollow cylinders\cite{Streubel14a, Streubel14, Streubel15a} or wires\cite{Nielsch01, Buchter13a, Rueffer12, Weber12} has become a dynamic research field. These 3D-shaped systems possess striking novel fundamental properties originating from the curvature-driven effects, such as magnetochiral effects\cite{Dietrich08, Otalora12a, Kravchuk12a, Hertel13a} and topologically induced magnetization patterns\cite{Smith11, Kravchuk12a, Pylypovskyi15b}. To this end, a general fully 3D approach was put forth recently to study dynamical and static properties of arbitrary curved magnetic shells and wires\cite{Gaididei14, Sheka15}. Due to the curvature and torsion in wires\cite{Sheka15} (Gaussian and mean curvatures in the case of shells\cite{Gaididei14}) two additional interaction terms appear in the exchange energy functional: a geometrically induced anisotropy term which is a bilinear form of the curvature and torsion, and an effective Dzyaloshinskii--Moriya interaction (DMI) term (Lifshitz invariants), which depends linearly on the curvature and torsion. In the framework of this approach, the existence of topologically induced patterns in Möbius rings\cite{Pylypovskyi15b} and new magnetochiral effects\cite{Gaididei14, Sheka15} were predicted.

In addition to these rich physics, the application potential of 3D-shaped objects is currently being explored as magnetic field sensorics for magnetofluidic applications\cite{Moench11, Muller12}, spin-wave filters\cite{Balhorn10, Balhorn12}, advanced magneto-encephalography devices for diagnosis of epilepsy at early stages\cite{Liu99a, Dumas13, Karnaushenko15a} or for energy-efficient racetrack memory devices\cite{Parkin08, Yan10}. The propagation of domain walls in a magnetic wire\cite{Catalan12} for racetrack memory\cite{Hayashi07, Parkin08} or magnetic domain wall logic \cite{Allwood02, Allwood05} applications induced by spin-polarized currents is already widely explored\cite{Vazquez15}. In contrast, spin-orbitronics \cite{Manchon14,Kuschel15}, based on current-induced spin-orbit torques, launches the new concept of low energy spintronic devices. 

Caused by the structural inversion symmetry, multilayers consisting of magnetic metal with nonmagnetic metal and oxide on contralateral sides like Pt/Co/Al$_x$O can support spin-orbit torques acting on the localized magnetic moments due to the Rashba and spin Hall effects\cite{Miron10,Martinez13}. The Rashba field, produced by a charge current in these structures is considered to be one of the 
}%
\twocolumn%
\noindent most efficient ways to act on the magnetization patterns\cite{Miron10}. However, in widely used planar devices, transverse domain walls are not affected by the Rashba effect\cite{Khvalkovskiy13}. Here, we demonstrate that the impact of the curvilinear effects on the magnetic texture of the domain walls in helical wires allows for their efficient displacement using spin-orbit Rashba torque. The geometrically induced anisotropy and DMI affect both the spatial orientation of the transverse (head-to-head and tail-to-tail) domain walls in helices as well as the magnetization
distribution in the domain wall. As a consequence, the chiral symmetry breaking is characteristic for the wall structure: the direction of the magnetization rotation in the wall is opposite for the left- and right-handed helices. The domain wall mobility is proportional to the product of curvature and torsion of the wire; it depends on the topological charge of the wall. The direction of the domain wall motion is determined by the sign of the product of the helix chirality and domain wall charge. Furthermore, a remarkable feature of this 3D geometry is that its curvature and torsion are coordinate independent. Therefore, all effects coupled with an interplay between the geometry of the system and the geometry of the magnetic texture may be presented here in a most clear and lucid style. The obtained results are general and valid for any thin wire with nonzero torsion.

\section*{Results}


We describe a helix curve by using its arc-length parametrization in terms of curvature--torsion:
\begin{equation} \label{eq:helix-param}
\vec{\gamma }(s) = \hat{\vec x} R\cos \left(\frac{s}{s_0}\right) + \hat{\vec y} R\sin \left(\frac{s}{s_0}\right) +\hat{\vec z} \frac{\mathcal{C}P s}{2\pi s_0},
\end{equation}
where $s$ is the arc length, $R$ is the helix radius, $P$ is the pitch of the helix, $\mathcal{C} = \pm 1$ is the helix chirality and $s_0 = \sqrt{R^2+ P^2/(2\pi)^2}$. A helix is characterized by the constant curvature $\kappa =R/s_0^2$ and torsion $\tau = \mathcal{C}P/(2\pi s_0^2)$. 

The magnetic properties are described using assumptions of classical ferromagnets with uniaxial anisotropy directed along the wire. The energy of the helix wire reads \cite{Sheka15c} 
\begin{equation*}
E = K^{\text{eff}} S \int  \mathscr{E}\mathrm{d}s, \qquad \mathscr{E} = \mathscr{E}^{\text{ex}} + \mathscr{E}^{\text{an}}.
\end{equation*}
Here $K^{\text{eff}} = K+\pi M_s^2$, where the positive parameter $K$ is a magnetocrystalline anisotropy constant of easy-tangential type, the term $\pi M_s^2$ stems from the magnetostatic contribution \cite{Slastikov12,Sheka15c,Yershov15b}, $M_s$ is the saturation magnetization, and $S$ is the cross-section area. The exchange energy density reads $\mathscr{E}^{\text{ex}} = -\ell^2\vec{m}\cdot \vec{\nabla}^2\vec{m}$, where $\vec{m}$ is the magnetization unit vector, $\ell =\sqrt{A/K^{\text{eff}}}$ is the characteristic magnetic length (domain wall width), and $A$ is an exchange constant. The anisotropy energy density is $\mathscr{E}^{\text{an}} = -\left(\vec{m} \cdot \vec{e}_{\text{an}}\right)^2$ where $\vec{e}_{\text{an}}$ is the unit vector along the anisotropy axis, which is assumed to be oriented along the tangential direction. The easy-tangential anisotropy in a curved magnet is spatially dependent. Therefore, it is convenient to represent the energy of the magnet in the curvilinear Frenet--Serret reference frame with $\vec{e}_{\textsc{t}}$ being a tangential (T), $\vec{e}_{\textsc{n}}$ being a normal (N) and $\vec{e}_{\textsc{b}}$ being a binormal (B) vector, respectively (TNB basis). 

\begin{figure}[t]
	\centering
	\includegraphics[width=\linewidth]{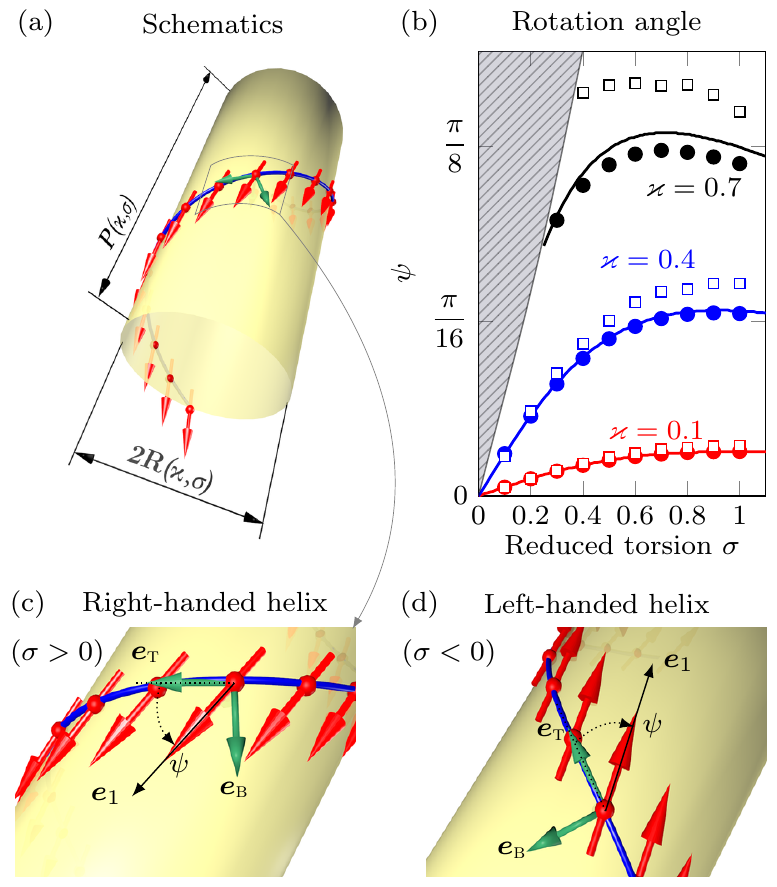}
	\caption{\textbf{Magnetization distribution in a helix:} (a)~Schematics of magnetic helix with the easy-tangential anisotropy (magnetic moments are shown with red arrows, TNB-basis is shown with green arrows. (b)~The rotation angle $\psi$ for different torsions $\sigma$ and curvatures $\varkappa$. The onion state is energetically preferable in the grey region~\cite{Sheka15c}. Solid lines correspond to analytics~\cite{Sheka15c}, filled circles and open squares correspond to \textsf{SLaSi} and \textsf{Nmag} simulations, respectively; see Methods for details. (c),~(d)~Discrete magnetic moments at equilibrium for right- and left-handed helices, respectively. The effective anisotropy axis is shown with thin black arrow~$\vec{e}_1$.
	}
	\label{fig:helix-tube}
\end{figure}

In the curvilinear frame, the exchange energy has three different contributions \cite{Sheka15}, $\mathscr{E}_{\text{ex}} = \mathscr{E}_{\mathrm{ex}}^{0} + \mathscr{E}_{\mathrm{ex}}^{\textsc{d}} + \mathscr{E}_{\mathrm{ex}}^{\textsc{a}}$. The first term $\mathscr{E}_{\mathrm{ex}}^{0} = \left| \vec{m}'\right|^2$, describes the isotropic part of the exchange expression, which has the same form as for a straight wire. Here and below the prime denotes the derivative with respect to the dimensionless coordinate $u  = s/\ell$. The second term, $\mathscr{E}_{\mathrm{ex}}^{\textsc{d}} = \mathcal{F}_{\alpha \beta } \left(m_\alpha  m_\beta ' - m_\alpha ' m_\beta  \right)$, is a curvature induced effective DMI, where the components of the Frenet--Serret tensor $\mathcal{F}_{\alpha \beta}$ are linear with respect to the reduced curvature and torsion 
\begin{equation*}
\varkappa = \kappa \ell,\qquad \sigma = \tau \ell,
\end{equation*}
respectively. The last term, $\mathscr{E}_{\mathrm{ex}}^{\textsc{a}} = \mathcal{K}_{\alpha \beta }m_\alpha m_\beta$, describes a geometrically induced effective anisotropy interaction, where the components of the tensor $\mathcal{K}_{\alpha \beta} = \mathcal{F}_{\alpha \nu} \mathcal{F}_{\beta \nu}$ are bilinear with respect to the curvature and torsion, see Supplementary Materials for details. Two additional contributions (effective DMI and effective anisotropy) naturally appear in the curvilinear reference frame similar to contributions to the kinetic energy of the mechanical particle in the rotating frame with  Coriolis force (linear with respect to velocity) and centrifugal force (bilinear with respect to velocity).

The emergent effective anisotropy leads to the modification of the equilibrium magnetic states\cite{Sheka15c}. Here, we consider helices with relatively small curvature possessing quasitangential magnetization distribution shown in Fig.~\ref{fig:helix-tube}(a). For further discussion it is instructive to project the magnetization onto the local rectifying surface, which coincides with the supporting surface of the helix [yellow cylinder in Fig.~\ref{fig:helix-tube}(a)]. The top view is plotted for the right-handed helix [$\sigma > 0$, Fig.~\ref{fig:helix-tube}(c)] and for the left-handed one [$\sigma <0$, Fig.~\ref{fig:helix-tube}(d)].

The influence of the curvature and torsion can be treated as an effective magnetic field ${F} \propto \sigma\varkappa$ acting along the binormal direction\cite{Sheka15}. This field causes a tilt of the the equilibrium magnetization from the tangential direction by an angle \cite{Sheka15c}:
\begin{equation*} 
\psi \approx \sigma \varkappa, \qquad \text{when} \quad \varkappa, |\sigma| \ll1,
\end{equation*}
see Fig.~\ref{fig:helix-tube}(b) and Supplementary Eq.~(S3) for details. The symbols represent the results of the spin-lattice simulations using the package \textsf{SLaSi} without magnetostatics and \textsf{Nmag} simulations of a magnetically soft wire, see Methods for details: the analysis shows that the model is adequate for soft magnets with $\varkappa \lesssim 0.4$.

Now we can rotate the reference frame in a local rectifying surface by the angle $\psi$ (see Supplementary for details). The magnetization in the rotated $\psi$-frame $\left\{\vec{e}_1, \vec{e}_3, \vec{e}_3\right\}$ reads
\begin{equation*} 
\vec{m} = \left(m_1,m_2,m_3\right) = \left(\cos\theta, \sin\theta \cos\phi, \sin\theta \sin\phi\right),
\end{equation*}
where magnetization angular variables $\theta$ and $\phi$ depend on  the spatial and temporal coordinates. Using this reference frame we can diagonalize the effective anisotropy energy density of the helix wire (Supplementary Eqs.~(S3)--(S5) for details):
\begin{equation} \label{eq:energy-helix}
\begin{split}
\mathscr{E} =& \underbrace{\left|\vec{m}'\right|^2}_{\text{isotropic exchange}}  \underbrace{-\mathscr{K}_1 m_1^2 + \mathscr{K}_2 m_2^2}_{\text{effective anisotropy}}\\
& \underbrace{+ \mathscr{D}_1 \left( m_2 m_3'-m_3 m_2'\right) + \mathscr{D}_2 \left( m_1 m_2'-m_2 m_1'\right)}_{\text{effective DMI}}.
\end{split}
\end{equation}
The coefficient $\mathscr{K}_1$ characterizes the strength of the effective easy-axis anisotropy  while $\mathscr{K}_2$ gives the strength of the effective easy-surface anisotropy. The parameters $\mathscr{D}_1$ and $\mathscr{D}_2$ are the effective DMI constants. We note that the energy \eqref{eq:energy-helix} has the general form of the energy density for 1D biaxial magnets with an intrinsic DMI and contains the complete set of the Lifshitz invariants. Hence, effective DMI constants $\mathscr{D}_1$ and $\mathscr{D}_2$ can include other contributions, e.\,g. the intrinsic DMI or DMI due to the structural inversion asymmetry\cite{Dzyaloshinsky58, Moriya60, Crepieux98}.

In the case of small curvature and torsion the geometrically induced anisotropy and DMI constants can be attributed to the geometrical parameters of the object:
\begin{equation*} 
\mathscr{K}_1\approx 1 + \sigma^2 - \varkappa^2, \quad \mathscr{K}_2\approx \varkappa ^2, \quad \mathscr{D}_1\approx 2\sigma, \quad \mathscr{D}_2\approx 2\varkappa.
\end{equation*}
The possible static magnetization structures can be found by variation of the total energy functional with density~\eqref{eq:energy-helix}. The homogeneous equilibrium state (quasitangential state) is described by $\theta^h = 0$ and $\theta^h = \pi$, which corresponds to the two possible directions of the helix magnetization.

\subsection*{Static domain wall}
\label{section:domain-wall}

One of the simplest inhomogeneous magnetization distribution in a nanowire is a transverse domain wall, which connects two possible equilibrium states. We start our analysis with general remarks about the domain wall described by the energy functional with the density~\eqref{eq:energy-helix}, which can be applied for a wide class of 1D magnets also with the intrinsic DMI.

The structure of the domain wall can be described analytically for $\mathscr{D}_2=\mathscr{K}_2=0$. This case corresponds to the uniaxial ferromagnet with an additional $\mathscr{D}_1$ DMI term. For such a system there is an exact analytical solution of static equations of the domain wall type:
\begin{equation} \label{eq:dw-ansatz}
\cos\theta ^{\text{dw}}(u) = -p\tanh \frac{u}{\delta},\qquad \phi ^{\text{dw}}(u) = \varPhi- \varUpsilon u.
\end{equation}
Here $p =\pm1$ is a domain wall topological charge: $p=1$ corresponds to kink (head-to-head domain wall) and $p=-1$ corresponds to antikink (tail-to-tail domain wall). The domain wall width $\delta$ and the slope $\varUpsilon$ are as follows:
\begin{equation} \label{eq:dw-parameters}
\delta  = \frac{1}{\sqrt{\mathscr{K}_1-\mathscr{D}_1^2/4}}, \qquad \varUpsilon = \mathscr{D}_1/2.
\end{equation}

In the uniaxial magnet with the anisotropy parameter $\mathscr{K}_1$ the typical domain wall width without DMI reads $\delta = 1/\sqrt{\mathscr{K}_1}$. One can see that the presence of DMI causes broadening of the wall. Furthermore, the domain wall is not perpendicular to the wire length and is titled by an angle determined by $\mathscr{D}_1$ constant. The slope of the azimuthal angle $\phi '=-\mathscr{D}_1/2$. This behaviour is similar to the known domain wall inclination in magnetic stripes caused by the intrinsic DMI\cite{Kravchuk14}.

\begin{figure*}[t]
	\centering
	\includegraphics[width=\linewidth]{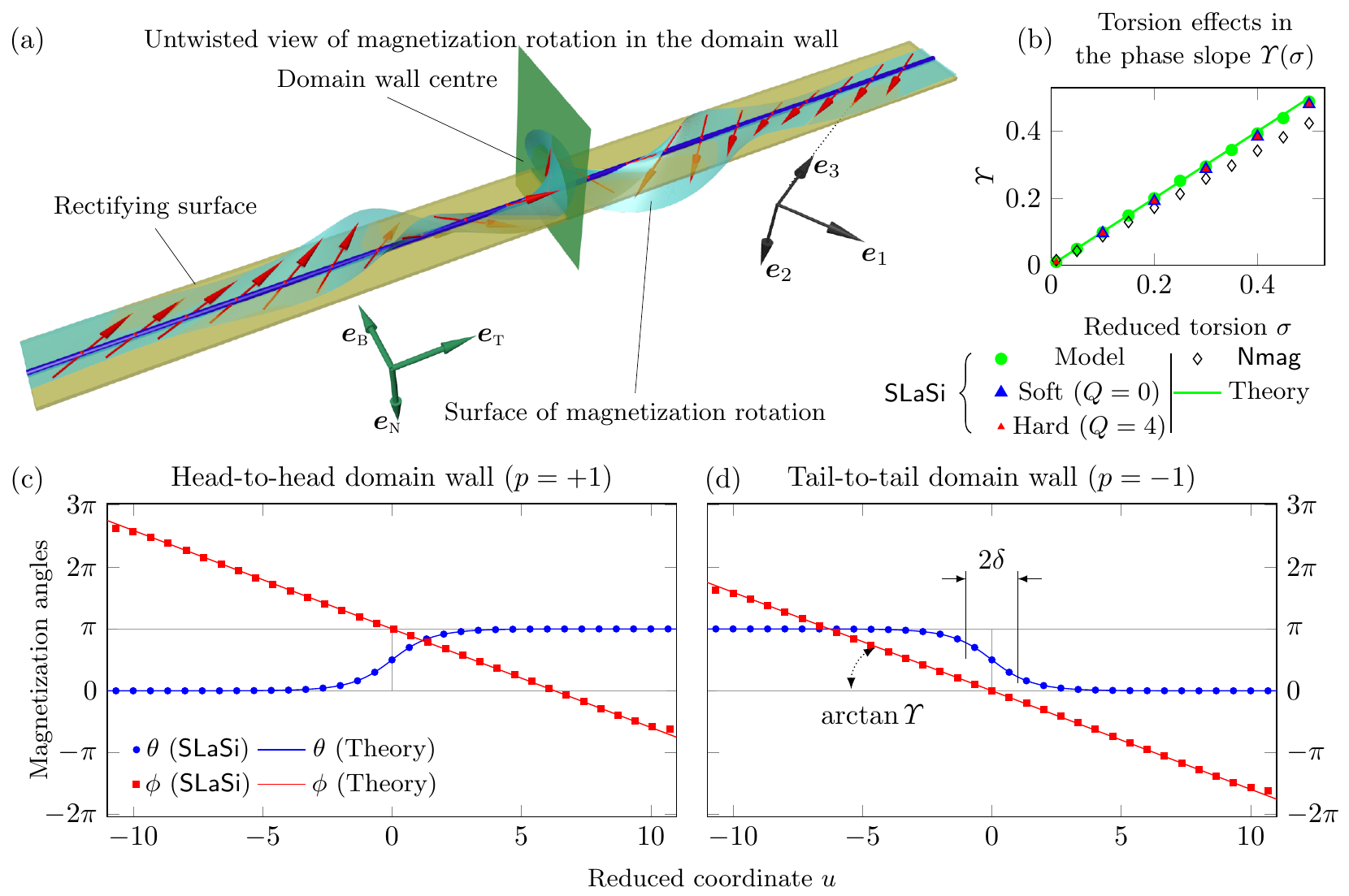}
	\caption{\textbf{Transverse domain walls in a helix:} (a)~Schematics of a domain wall in the helix ($\sigma > 0$), untwisted view. Magnetic moments (red arrows) lie on the helix wire (blue cylinder), directed along $\vec{e}_\textsc{t}$. Magnetic moments inside domains are parallel to $\vec{e}_1$. (b)~Phase slope $\varUpsilon(\sigma)$ for $\varkappa = 0.1$ [symbols correspond to simulations and solid line is accordingly to Eq.~\eqref{eq:phaseSlope}]. Symbols represent the results of the \textsf{SLaSi} simulations: for anisotropic wire without magnetostatics (model, green circle), magnetically soft wire (blue triangle) and magnetically hard wire (open triangle). Diamonds correspond to the micromagnetic simulations of a magnetically soft sample performed using \textsf{Nmag}, see Methods for details. (c),~(d)~Magnetization angles in the $\psi$-frame [black arrows in panel~(a)] for the head-to-head and tail-to-tail domain walls, respectively; $\varkappa = 0.1$, $\sigma = 0.5$. Symbols correspond to simulations (each tenth chain site is plotted), and solid lines to Ansatz~\eqref{eq:dw-ansatz}. Thin grey lines show levels 0, $\pi$ and centre of the domain wall. 
	}
	\label{fig:untwistedDWall}
\end{figure*}

In the following we proceed with the investigation of the finite curvature effects on the magnetization distribution in domain walls in helices.  We will apply a variational approach by using \eqref{eq:dw-ansatz} as a domain wall Ansatz with the domain wall width $\delta$, initial phase $\varPhi$, and the slope $\varUpsilon$ being the variational parameters. By inserting Eq. \eqref{eq:dw-ansatz} into the energy density functional \eqref{eq:energy-helix} and integrating over the arclength variable $s$, we obtain
\begin{equation*} 
\begin{split}
\frac{E^{\text{dw}}}{K^\text{eff} S \ell} = & \underbrace{\dfrac{2}{\delta}  + 2\delta \varUpsilon^2}_{\text{exchange}}  \underbrace{+2\delta\mathscr{K}_1 + \delta \mathscr{K}_2 \left(1+\mathscr{C}_1 \cos2{\varPhi}\right) }_{\text{effective anisotropy}}\\
&  \underbrace{- 2\delta \mathscr{D}_1 \varUpsilon + p \mathscr{C}_2 \mathscr{D}_2 \cos{\varPhi}}_{\text{effective DMI}}, \\
\mathscr{C}_1&=  \frac{\pi \delta \varUpsilon}{\sinh (\pi \delta \varUpsilon)}, \qquad \mathscr{C}_2 = \frac{\pi   (1 + \delta^2 \varUpsilon^2)}{ \cosh (\pi \delta \varUpsilon/2)}.
\end{split}
\end{equation*}

The presence of the effective DMI with the constant $\mathscr{D}_2$ breaks the symmetry of the domain walls with opposite topological charges $p$, which is coupled with the domain wall phase ${\varPhi}$: for the small enough torsion and curvature the energetically preferable domain wall with the topological charge $p$ has the equilibrium phase ${\varPhi} = (1+p)\pi/2$. In the case $\varkappa,|\sigma|\ll1$,  one can find 
\begin{equation} \label{eq:phaseSlope}
\varUpsilon \approx \sigma \qquad \text{and} \qquad  \delta \approx 1.
\end{equation}
The variational parameters~\eqref{eq:phaseSlope} coincide with parameters~\eqref{eq:dw-parameters} of the exact solution obtained in the case $\mathscr{D}_2=\mathscr{K}_2=0$. Thus, the approximation of vanishing curvatures describe the domain wall statics for small enough $\varkappa$ and $\sigma$.

\begin{figure*}[t!]
	\centering
	\includegraphics[width=\linewidth]{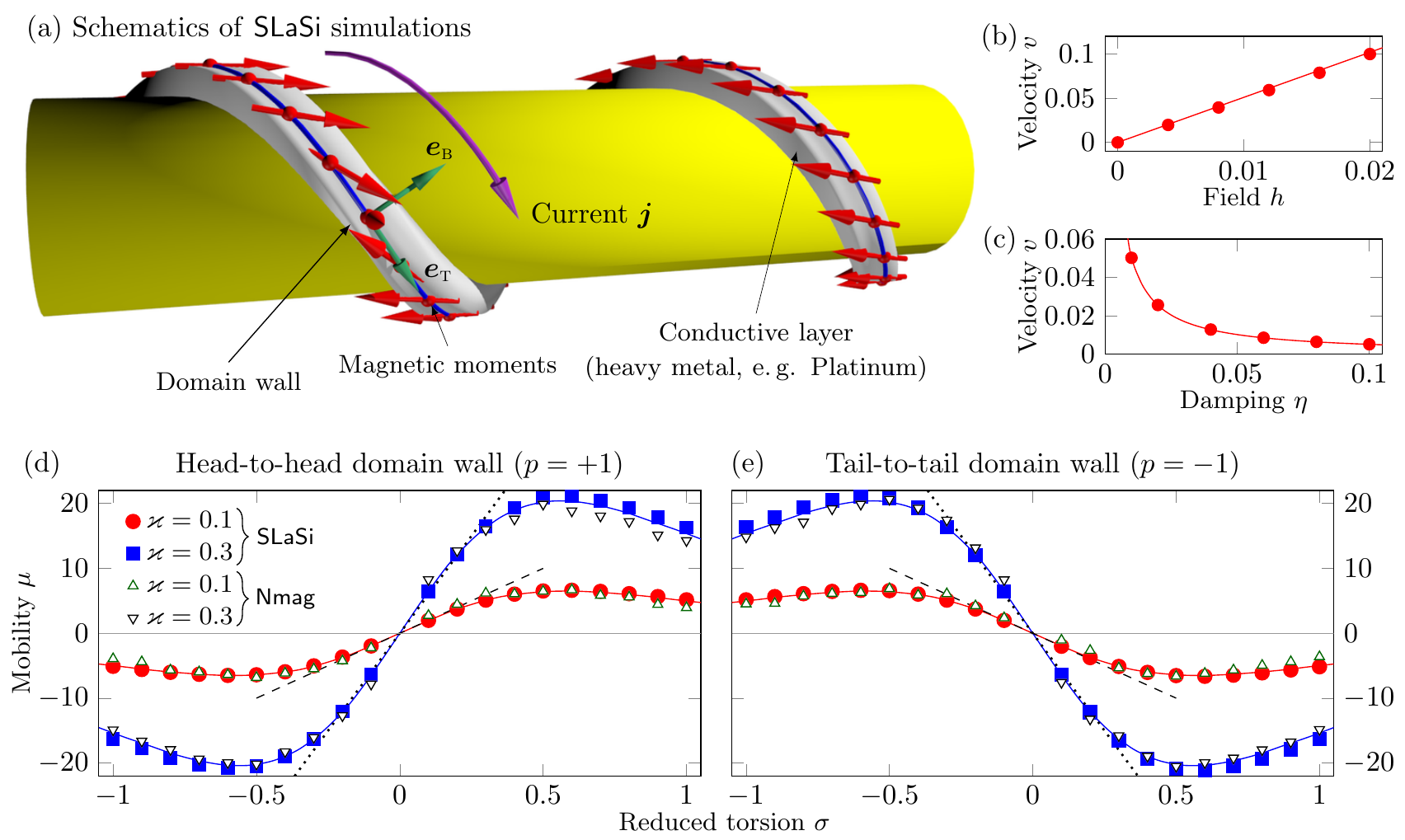}
	\caption{\textbf{Domain wall motion by the Rashba spin-orbit torque:} Symbols correspond to simulations and solid lines are calculated accordingly to Eq.~\eqref{eq:velocity}. (a)~Schematics of the domain wall dynamics: magnetic moments (red arrows) lie on a conductive wire (grey) (direction of the current $\vec{j}$ along $\vec{e}_\textsc{t}$ is shown with magenta arrow). The Rashba field $\vec{H}^{\textsc{r}}$ acts along $\vec{e}_{\textsc{b}}$. (b), (c)~Wall velocity as a function of the applied field and damping for $\varkappa=0.1$ and $\sigma = 0.3$. The mobility of the head-to-head (d) and tail-to-tail (e) domain walls in weak fields as a function of the reduced torsion. Dashed and dotted lines show asymptotics~\eqref{eq:velocity-weak-sigma} for $\varkappa = 0.1$ and $\varkappa = 0.3$, respectively. \textsf{Nmag} simulations of magnetically soft samples with $\varkappa = 0.1$ and different torsions, $\sigma=0.1$ (f) and $\sigma = -0.1$ (g). Under the action of the electric current $\vec{j}$ domain walls move in the opposite directions starting from the central position.}
	\label{fig:mobility}
\end{figure*}

The comparison of these predictions with the 3D spin-lattice simulations using package  \textsf{SLaSi}\cite{slasi}, and micromagnetic simulations using \textsf{Nmag}\cite{Fischbacher07} confirms our theory, see Fig.~\ref{fig:untwistedDWall} (the details of simulations are described in Methods). Figure~\ref{fig:untwistedDWall}(a) represents the untwisted view of the domain wall. The magnetization direction corresponds to the ground state along $\vec{e}_1$ inside two domains. Inside the head-to-head domain wall the magnetization is directed outward the helix (opposite to $\vec{e}_{\textsc{n}}$). Qualitatively this is explained by the fact that such a configuration minimizes the magnetization gradient and, therefore, the exchange energy. For the tail-to-tail domain wall the direction of the magnetization tilt is opposite to the head-to-tail one. The dependence of the phase slope $\varUpsilon$~\eqref{eq:phaseSlope} on the torsion $\sigma$ is in good agreement with the simulation data, solid line in Fig.~\ref{fig:untwistedDWall}(b). Symbols correspond to the results of the simulations carried out for $\varkappa = 0.1$. We performed the spin-lattice simulations without magnetostatics (green circles) and with magnetostatics for for a magnetically soft sample (blue filled triangles, the quality factor $Q = K/2\pi M_\textsc{s}^2$ equals to zero) as well as magnetically hard sample with $Q = 4$. The micromagnetic simulations of a thin Permalloy wire (diamonds) are also in good agreement with the spin-lattice simulations and theory. The static head-to-head and tail-to-tail domain walls are well described by the Ansatz~\eqref{eq:dw-ansatz} with optimal parameters determined by \eqref{eq:phaseSlope}, see solid lines in Fig.~\ref{fig:untwistedDWall}(c),~(d), for $\varkappa = 0.1$, $\sigma=0.5$.

\subsection*{Domain wall dynamics driven by the Rashba spin-orbit torque}
\label{sec:dynamics}

\begin{figure*}[t!]
	\centering
	\includegraphics[width=\linewidth]{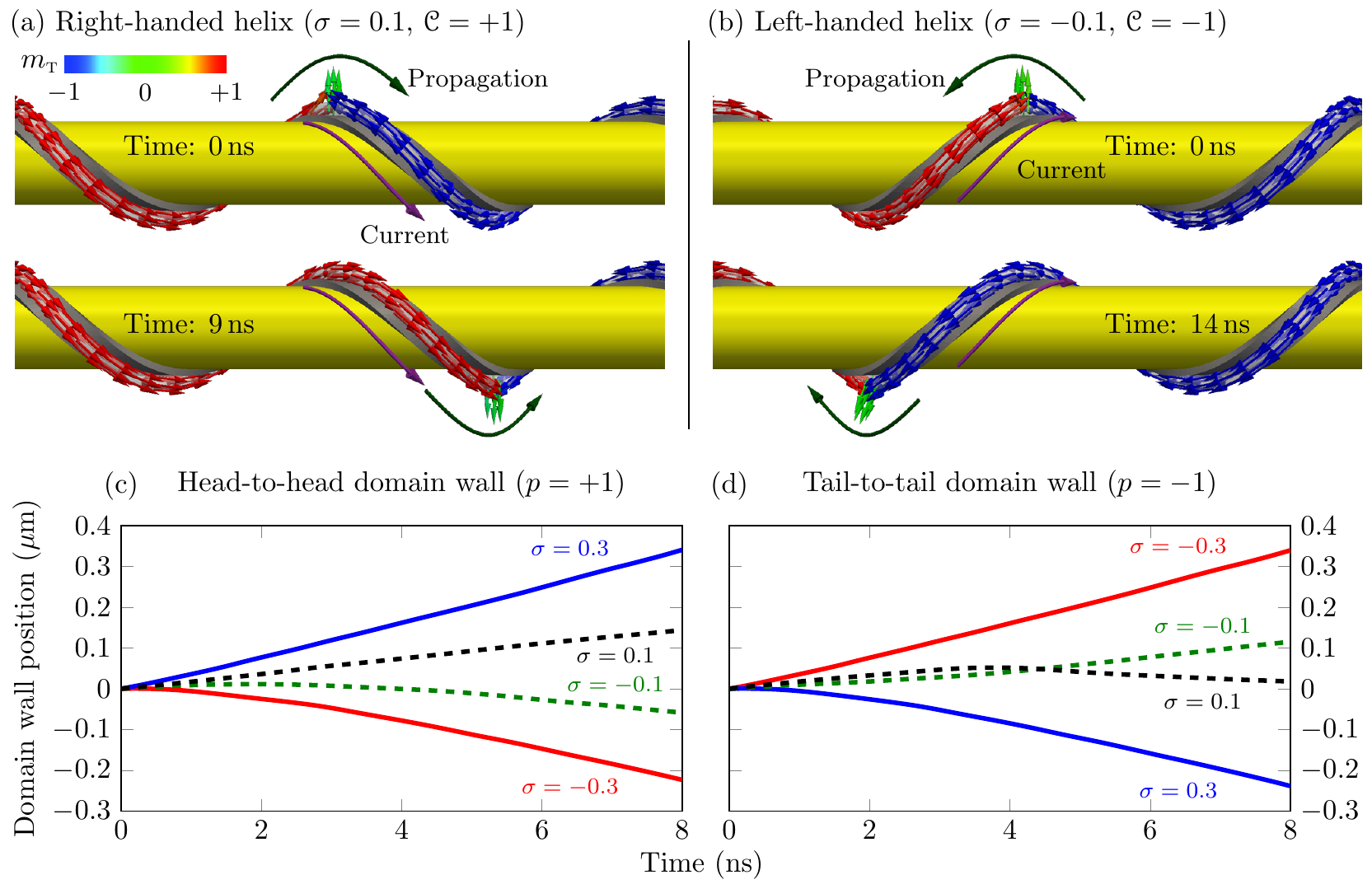}
	\caption{\textbf{\textsf{Nmag} simulations of the domain wall motion in a helix with $\varkappa=0.1$:} Head-to-head domain wall ($p = 1$) in helices with $\sigma=0.1$ (a) and $\sigma = -0.1$ (b) under the action of the Rashba field $h = 0.02$ (using SI units $H^\textsc{r}\approx 10.8$\,mT). The direction of the electric current (along $\vec{e}_\textsc{t}$) and domain wall motion are shown with violet and dark-green arrows, respectively. Time behaviour of the domain wall position for head-to-head (c) and tail-to-tail (d) domain walls in helices with $\varkappa = 0.1$, see also Supplementary Video~\ref{mov:motion}. All curves are matched at zero time and coordinate.}
	\label{fig:dwmotion}
\end{figure*}

Here, we  describe the domain wall dynamics in the Rashba spin-orbit system \cite{Obata08}, where the magnetic wire is adjacent to a nonmagnetic conductive layer with a strong spin-orbit interaction. The Rashba effect typically appears in systems with inversion symmetry broken spin-orbit interaction \cite{Manchon09}. We consider the parallel geometry, in which the ferromagnetic wire is parallel to the spin-orbit layer on the whole length of the wire~\cite{Khvalkovskiy13}.

The sketch of the system is shown in Fig.~\ref{fig:mobility}(a). The magnetic wire is winded around the conductive layer forming a helix. The electrical charge current $\vec{j}$ flows along the magnetic wire in the tangential direction $\vec{e}_{\textsc{t}}$. Under the action of the field-like torque caused by the Rashba effect, the magnetic subsystem is affected by the effective Rashba field \cite{Khvalkovskiy13}
\begin{equation*} 
\vec{H}^{\textsc{r}} = \frac{\alpha \mathcal{P}}{\mu _B M_s} \left[\vec{j}\times \vec{n}\right]
\end{equation*}
with $\alpha$ being the Rashba parameter, $\mathcal{P}$ being the polarization of the carriers in the ferromagnetic layer, $\mu _B$ being the Bohr magneton and $\vec{n}$ being the unit vector perpendicular to the spin-orbit layer. 

In such parallel geometry the Rashba field is always directed perpendicular to the wire. For a straight wire the direction of the Rashba field is transversal to the domain magnetization, hence the field can not push the wall \cite{Khvalkovskiy13}. However for the helix geometry the equilibrium magnetization direction deviates from the wire direction. The energy density of the interaction with the effective Rashba field is $\mathscr{E}^{\textsc{r}} =  - 2\vec{h}\cdot\vec{m}$, where $\vec{h} = \vec{H}^{\textsc{r}}/H^{\textsc{a}}$ is the reduced field normalized by the anisotropy field $H^{\textsc{a}} = 2K^{\text{eff}}/M_s$. There are two components of the magnetic field: $h_\|  = h\sin\psi$ is parallel along the domain, hence it pushes the wall. Another one, $h_\perp = h\cos\psi$ is directed along $\vec e_2$. In general, magnetic fields with the transversal component results in the deformation of the domain wall profile  and other changes of the characteristic parameters like Walker field and maximal domain wall velocities~\cite{Sobolev95, Bryan08, Lu10, Goussev13b}. However, in the case of weak fields, we can limit our consideration to the parallel field $h_\|$ only and neglect the dynamical changes of the wall width. Furthermore, we will not take into account the influence of {\O}rsted fields generated by the charge current. 

Far below the Walker limit, we can use the generalized $q$--$\varPhi$ model \cite{Kravchuk14}, cf.~\eqref{eq:dw-ansatz}:
\begin{equation} \label{eq:dw-ansatz-dynamics}
\begin{split}
\cos\theta ^{\text{dw}}(u,\bar{t}) &= -p\tanh \frac{u-q(\bar{t})}{\delta},\\
\phi ^{\text{dw}}(u,\bar{t})   &= \varPhi(\bar{t})-\varUpsilon \left[u-q(\bar{t})\right],
\end{split}
\end{equation}
where $\bar{t}= \omega_0 t$ and $\omega_0 = \upgamma_e K^{\text{eff}}/M_s$, $\upgamma_e$ being the gyromagnetic ratio.

Using $(q,\varPhi)$ as a pair of time dependent collective coordinates, we obtain the stationary motion of the domain wall (see Methods for details)
\begin{equation} \label{eq:velocity}
v \equiv \frac{\mathrm{d}q}{\mathrm{d}\bar{t}} (\bar{t}\to \infty) = \frac{2 p h \delta }{\eta} \cdot \dfrac{\sin\psi}{1+\delta^2\varUpsilon^2}.
\end{equation}

We checked the theoretically predicted velocities for the domain wall motion \eqref{eq:velocity} by \textsf{SLaSi} and \textsf{Nmag} simulations in the range of effective fields, $h = 0\div 0.02$, see Figs.~\ref{fig:mobility}(b)--(d) and Methods for details. Symbols correspond to \textsf{SLaSi} and \textsf{Nmag} simulations, solid lines correspond to the theoretical predictions, obtained accordingly to Eq.~\eqref{eq:velocity}, see also Supplementary Eq.~(S3). The domain wall velocity is almost linear with the field, see  Fig.~\ref{fig:mobility}(b) [with a fixed damping constant $\eta = 0.1$]. The inverse linear dependence $v \propto 1/\eta$ is well pronounced in Fig.~\ref{fig:mobility}(c). The maximal velocity $v = 0.1$ shown in Fig.~\ref{fig:mobility}(b) for $h=0.02$ corresponds to $35$~m/s for Permalloy.

The most intriguing effect in the domain wall dynamics is the torsion dependence of the wall motion. The mobility of the domain wall $\mu = v/ h$ as a function of the helix torsion is plotted in Fig.~\ref{fig:mobility}(d),~(e) for different helix curvatures. In the case of small curvature and torsion ($\varkappa, |\sigma| \ll1$), the wall mobility, accordingly to \eqref{eq:velocity}, has the following asymptotic: 
\begin{equation} \label{eq:velocity-weak-sigma}
\mu \approx \frac{2 p \delta}{\eta}\cdot \varkappa \sigma.
\end{equation}
Therefore, the domain wall can move only under the joint action of the curvature and torsion. The direction of the domain wall motion depends on the helix chirality $\mathcal{C}$, see Fig.~\ref{fig:dwmotion}(a),~(b), where the head-to-head domain wall position is shown at different time moments and Fig.~\ref{fig:dwmotion}(c),~(d), where the domain wall position is shown as a function of time for different torsions and values of $p$. The initial domain wall displacement occurs in the positive direction, while the steady-state motion is described by Eq.~\eqref{eq:velocity}. That is why the close positions of the domain walls in Fig.~\ref{fig:dwmotion}(a) and~(b) occur at different time of 9\,ns and 14\,ns.

In some respect, the effect of chirality sensitive domain wall mobility is similar to the recently found chiral-induced spin selectivity effect\cite{Goehler11,Naaman12} in helical molecules due to the Rashba interaction \cite{Eremko13}.

\section*{Discussion}

First, we discuss the consequence of the interplay between the curved geometry of the helical wire with the magnetic texture of the transverse domain walls:

(i) The geometrically induced effective anisotropy causes the tilt of the equilibrium magnetization by the angle $\psi$ with respect to the tangential direction. This rotation angle depends on the product of the curvature and the torsion. There appears curvature induced easy-surface anisotropy. For the helix geometry the anisotropy tends to orient the magnetization within the rectifying surface, i.\,e. tangentially to the cylinder surface. Additionally, the geometry caused easy-axis anisotropy, favours the orientation of the magnetization along $\vec{e}_1$ direction.

(ii) The more intriguing features of the geometry are connected to the curvatures induced Dzyaloshinskii--Moriya interaction. Two effective DMI terms in the energy \eqref{eq:energy-helix} correspond to all possible Lifshitz invariants in the 1D case. In this respect our analysis is valid also for 1D systems with an intrinsic DMI as well as for the DMI induced due to the structural inversion asymmetry. Using SI units, one estimates that $D_1\approx 4\pi A P/(R^2+P^2)$. Using typical values $A=10$ pJ/m, we obtain that $D_1=0.28$ mJ/m$^2$ for a helix with the radius $R=50$ nm and the pitch $P=300$ nm; $D_1=0.14$ mJ/m$^2$ for $R=100$ nm,  $P=600$ nm. These values are comparable to those estimated from the \textit{ab initio} calculations for multilayer systems\cite{Grigoriev07,
	Yang15}. 

It is instructive to compare the geometrically induced DMI in helices with the intrinsic DMI for the untwisted objects. In this work we restricted ourselves by considering the quasitangential ground state of the helix, which is realized for the relatively weak curvatures and torsions (weak effective DMI)~\cite{Sheka15c}. In case of strong DMI, the helix favours the onion ground state~\cite{Sheka15c}, where the magnetization is almost homogeneous (in the physical space) due to the strong exchange interaction. At the same time, the magnetization rotates in the curvilinear reference frame. Such a state is an analogue of the spiral state in straight magnets with intrinsic DMI.

\begin{figure*}[t!]
	\centering
	\includegraphics[width=\linewidth]{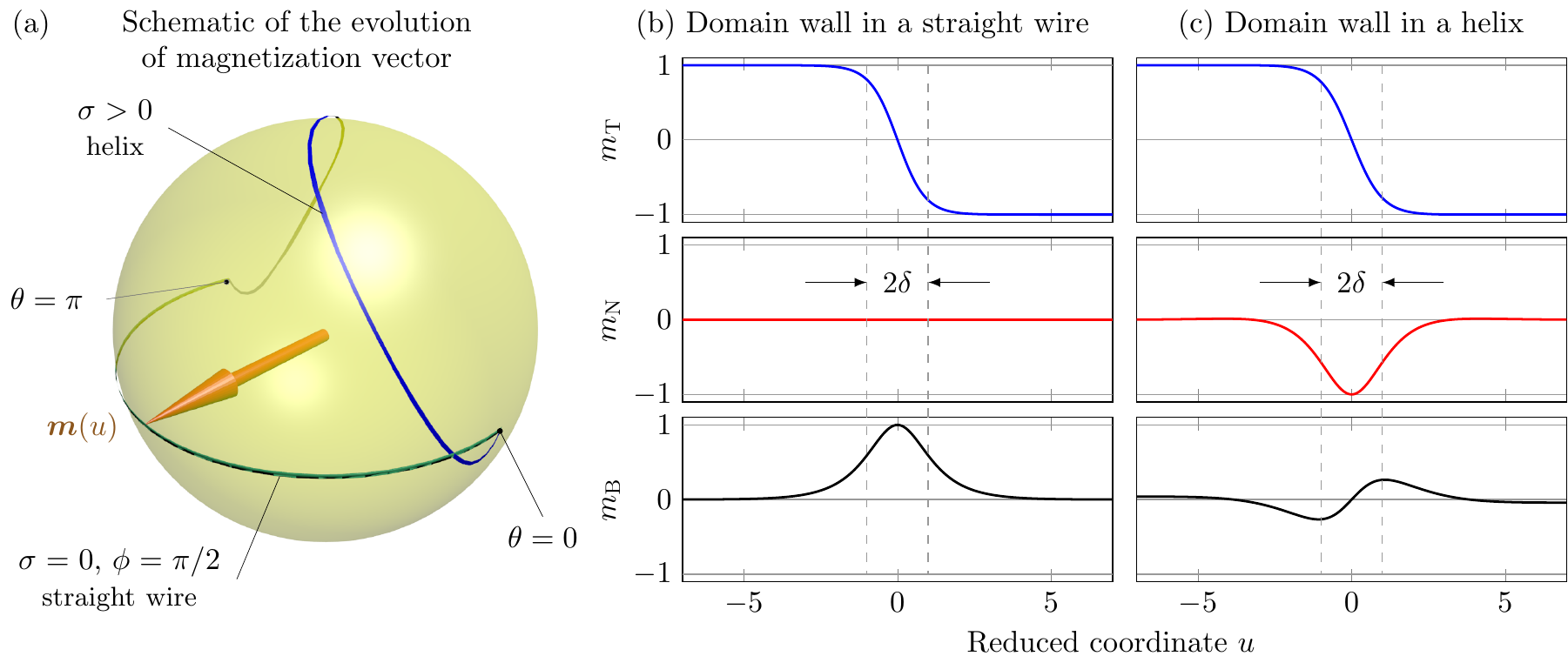}
	\caption{\textbf{The role of the curvature induced DMI: comparison of the magnetization distribution in a helix and a biaxial straight wire.} (a) The evolution of the magnetization vector $\vec{m}(u)$ on a unit sphere for a domain wall in a helix ($\sigma > 0)$ and straight wire ($\sigma = 0$). (b), (c) Tangential $m_\textsc{t}$, normal $m_\textsc{n}$ and binormal $m_\textsc{b}$ magnetization components of the domain wall in a straight wire and a helix: while $m_\textsc{t}$ have the similar shape, other components are different due to appearance of the effective DMI.
	}
	\label{fig:comparison00}
\end{figure*}
(iii) The geometrically induced DMI drastically changes the internal structure of the transverse domain wall: the azimuthal magnetization angle $\phi$ rotates inside the wall, see Supplementary Fig.~S1. While the domain wall orientation in its centre is determined by the domain wall topological charge $p$, the direction of the magnetization rotation (i.\,e. magnetochirality $\mathfrak{C} = -\text{sgn}\,\varUpsilon = - \text{sgn}\,\sigma$) mainly depends on the helix torsion $\sigma$. One can interpret the sign of $\sigma$ as the helix chirality $\mathcal{C}$ (different for right-handed helix when $\sigma > 0$ and left-handed one when $\sigma < 0$). Therefore, the magnetochirality of the domain wall is always opposite to the helix chirality, $\mathfrak{C} = - \mathcal{C}$.

In order to elucidate the role of the geometrically induced DMI we compare the domain wall structure in a helix with the domain wall in a straight wire of a biaxial magnet without DMI. Figure~\ref{fig:comparison00} shows the comparison of the magnetization distribution for these two geometries obtained by the \textsf{SLaSi} simulations. The panel (b) represents the data for a straight wire with the energy~\eqref{eq:energy-helix}, where the anisotropy coefficients $\mathscr{K}_1$, $\mathscr{K}_2$ correspond to the effective anisotropies in the helix, and DMI constants $\mathscr{D}_1 = \mathscr{D}_2 = 0$. The panel (c) represents the data for a helix with $\varkappa = 0.1$, $\sigma = 0.5$. While for the straight wire the magnetization always lies in the plane, $m_{\textsc{n}} = 0$, the competition between the easy-plane anisotropy and DMI results in the essential coordinate dependence of both normal and binormal magnetization components.

(iv) The chiral symmetry breaking strongly impacts the domain wall dynamics and allows the motion of domain walls under the action of the Rashba spin-orbit torque: the direction of motion if determined by the product of the helix chirality and the wall charge ($v\propto \sigma p$). The wall does not move in the limit of a planar wire, see Fig.~\ref{fig:mobility}. The head-to-head and tail-to-tail domain walls move in opposite directions, see Supplementary Video~\ref{mov:motion}. Our theory describes the domain wall motion both in magnetically hard and soft helices, see comparison in Fig.~\ref{fig:untwistedDWall}(b) for the phase slope and~\ref{fig:mobility}(b),~(c) for the domain wall mobility, and also Supplementary Fig.~S2.
The results obtained for this test system are valid well beyond the considered here specific case of helical wires. 
The Rashba torque driven domain wall motion will be characteristic for any transverse wall present in a curvilinear system with non-zero torsion.

\section*{Methods}
\label{sec:methods}

\begin{methods}
	
	\subsection*{Spin-lattice and micromagnetic simulations}
	
	Numerically we study the magnetization textures in a helix and its dynamics using the in-house developed spin-lattice simulator \textsf{SLaSi}~\cite{slasi} for anisotropic samples and \textsf{Nmag}\cite{Fischbacher07} for magnetically soft samples. 
	
	When using \textsf{SLaSi} we consider a classical chain of magnetic moments $\vec{m}_i$, with $i = \overline{1,N}$, situated on a
	helix \eqref{eq:helix-param}. We use the anisotropic Heisenberg Hamiltonian taking into account the exchange interaction, easy-tangential anisotropy and Rashba field. The dynamics of this system is described by a set of $N$ vector Landau-Lifshitz ordinary differential equations, see Ref. \cite{Pylypovskyi14} for the general description of the \textsf{SLaSi} simulator and Ref. \cite{Sheka15c} for details of the helix simulations. To study the static magnetization distribution spin chains of $N=2000$ sites are considered. The domain wall is placed in the centre of the chain. To simulate the magnetization dynamics spin chains of 4000~sites are considered. The domain wall is placed at the 300-th site from one end of the helix and is pushed by the field-like torque to another end. The velocity is measured at the steady state of the domain wall motion before it is driven out off the helix. In all simulations the magnetic length $\ell = 15a$ with $a$ being the lattice constant and damping $\eta = 0.01$ is used except the case when studying the velocity dependence on damping, where $\eta = 0.01\ldots 0.1$. For all simulations with magnetostatics the exchange length $\ell_\text{ex}$ is used to obtain the effective magnetic length $\ell = \sqrt{A/K^\text{eff}} = 2\ell_\text{ex}/\sqrt{1+2Q} = 15a$.
	
	The simulations using the \textsf{Nmag} are performed with the following parameters: exchange constant $A = 13$\,pJ/m, saturation magnetization $M_\textsc{s} = 860$\,kA/m and damping $\eta = 0.01$ which correspond to Permalloy (Ni$_{81}$Fe$_{19}$). These parameters result in the effective anisotropy field of $H^\textsc{A}_\text{Py} = 0.54$\,T and exchange length $\ell_\text{ex} \approx 3.7$\,nm. Samples of radius 5\,nm and length 1\,$\upmu$m are studied. Thermal effects and anisotropy are neglected. The typical Rashba field $h = 0.02$ (using SI units $H^\textsc{r} \approx 10.8$\,mT) corresponds to the electrical charge current density $j = 10.8$\,mA/$\upmu$m$^2$ for the polarization of carriers $\mathcal{P} = 0.5$ and Rashba parameter $\alpha = 100$\,peV\,m\cite{Miron10}. The static and dynamical properties of the domain walls on a helix are studied in the same way as for the classical chain described above.
	
	The simulations are performed using the computer clusters of the Bayreuth University\cite{btrzx}, Taras Shevchenko National University of Kyiv\cite{unicc} Bogolyubov Institute for Theoretical Physics of the National Academy of Sciences of Ukraine\cite{bitpcluster}.
	
	\subsection*{Domain wall dynamics}
	
	We use the generalized collective coordinate $q$--$\varPhi$ approach \cite{Kravchuk14} based on the effective Lagrangian formalism. Inserting the Ansatz \eqref{eq:dw-ansatz-dynamics} into the ``microscopic'' Lagrangian with the density $\mathscr{L} = -\cos\theta\dot{\phi} - \mathscr{E}$ and the dissipative function $\mathscr{F} = \frac{\eta}{2} \left[{\dot{\theta}}^2 + \sin^2\theta {\dot{\phi}}^2\right]$, after integration over the wire, we obtain the effective Lagrangian and the effective dissipative function, normalized by $K^{\text{eff}}S\ell$, as follows:
	\begin{equation} \label{eq:L-eff}
	\tag{M1} %
	\begin{split}
	L^{\text{eff}} &= G^{\text{eff}}-E^{\text{eff}},\qquad G^{\text{eff}} = 2p\varPhi\dot{q},\\
	E^{\text{eff}} &=\dfrac{2}{\delta}  +\delta\left[2\mathscr{K}_1  + 2 \varUpsilon^2 + \mathscr{K}_2 \left(1+\mathscr{C}_1 \cos2\varPhi\right)\right]
	- 2\delta \mathscr{D}_1 \varUpsilon\\
	& + p \mathscr{C}_2 \mathscr{D}_2 \cos\varPhi - 4 p h q \sin\psi, \qquad F^{\text{eff}} = \eta \left[\frac{{\dot q}^2}{\delta} + \delta \left(\dot{\varPhi} + \varUpsilon \dot{q}\right)^2\right].
	\end{split}
	\end{equation}
	Here and below overdot means the derivative over $\bar{t}$. The effective equations of motion are then obtained as the Euler–Lagrange-Rayleigh equations
	\begin{equation} \label{eq:ELR}
	\tag{M2} %
	\frac{\partial L^{\text{eff}}}{\partial X_i} - \frac{\mathrm{d}}{\mathrm{d}\bar{t}} \frac{\partial L^{\text{eff}}}{\partial \dot{X}_i} = \frac{\partial F^{\text{eff}}}{\partial \dot{X}_i}, \qquad X_i = \left\{q, \varPhi \right\}.
	\end{equation}
	These equation describe the steady motion of the domain wall $q=q_0 + v \bar{t}$ with the constant velocity \eqref{eq:velocity}. The corresponding phase $\varPhi =\text{const}$ is determined by the equation $2\mathscr{C}_1 \mathscr{K}_2 \delta \sin2\varPhi + p \mathscr{C}_2 \mathscr{D}_2\sin\varPhi = -2v \left(p-\delta \eta \varUpsilon \right)$.
	
	\section*{Acknowledgements}
	
	O.~P. acknowledges a financial support from DAAD (Code No. 91530902-FSK). O.~P. and D.~S. thank F.~G.~Mertens for helpful discussions, thank the University of Bayreuth, where part of this work was performed, for kind  hospitality. O.~P., D.~S. and V.~Kr. acknowledge the support from the Alexander von Humboldt Foundation. This work is financed in part via the ERC within the EU Seventh Framework Programme (ERC Grant No.~306277) and the EU FET Programme (FET-Open Grant No.~618083).
	
	\section*{Author contributions statement}
	
	O.~P. and D.~Sh. formulated the theoretical problem and performed the analytical calculations. O.~P. performed spin-lattice simulations. K.~Yer. performed micromagnetic simulations. O.~P., D.~Sh., V.~Kr., K.~Yer., D.~M. and Yu.~G. contributed to the discussion and writing of the manuscript text.
	
	\section*{Additional information}
	
	\textbf{Competing financial interests:} The authors declare no competing financial interests.

\end{methods}



\renewcommand{\thefigure}{\thefigurea}
\setcounter{figure}{0}

\newtagform{suppl}[]{(S}{)}
\makeatletter
\renewcommand{\thefigure}{S\@arabic\c@figure}
\makeatother

\appendix
\twocolumn[%
\textbf{ \Large{\large Supplementary to}\\[2mm]
	Rashba Torque Driven Domain Wall Motion in Magnetic Helices\\[5mm]
}]

\usetagform{suppl}
%
%
%
\section{The model}

Let us consider a curvilinear magnetic wire, which can be modelled by the 3D curved $\vec{\gamma }\subset \mathbb{R}^3$. We describe the magnetic properties of the wire using assumptions of classical ferromagnet with uniaxial anisotropy directed along the wire. The easy-tangential anisotropy in a curved magnet is spatially dependent. In order to describe the magnetization distribution in such systems it is convenient to use a curvilinear Frenet--Serret (TNB) parametrization of the curve $\vec{\gamma }$:
\begin{equation*} 
\vec{e}_{\textsc{t}} = \partial_s\vec{\gamma }, \qquad \vec{e}_{\textsc{n}} = \frac{\partial_s\vec{e}_{\textsc{t}}}{\left|\partial_s\vec{e}_{\textsc{t}} \right|}, \qquad \vec{e}_{\textsc{b}} = \vec{e}_{\textsc{t}}\times \vec{e}_{\textsc{n}}
\end{equation*}
with $\vec{e}_{\textsc{t}}$ being the tangent, $\vec{e}_{\textsc{n}}$ being the normal, and $\vec{e}_{\textsc{b}}$ being the binormal to $\vec{\gamma }$ and $s$ being the arc length. In particular, we use TNB parametrization of the magnetization unit vector,
\begin{equation} \label{eq:m-TNB}
\vec{m} = \begin{pmatrix} m_{\textsc{t}}, m_{\textsc{n}}, m_{\textsc{b}} \end{pmatrix}^T
\end{equation}
with the curvilinear components $m_\alpha $. Here and below Greek indices $\alpha ,\beta $ numerate curvilinear coordinates (TNB-coordinates) and curvilinear components of vector fields. For an arbitrary thin wire the energy can be presented as follows \cite{Sheka15}
\begin{equation} \label{eq:energy}
\begin{split}
E &= K^{\text{eff}}S\,\int\! \mathscr{E}\mathrm{d}s, \qquad \mathscr{E} = \mathscr{E}_{\text{ex}} + \mathscr{E}_{\text{an}},\\
\mathscr{E}_{\text{ex}} &= \mathscr{E}_{\mathrm{ex}}^{0} + \mathscr{E}_{\mathrm{ex}}^{\textsc{d}} + \mathscr{E}_{\mathrm{ex}}^{\textsc{a}} , \quad \mathscr{E}_{\mathrm{ex}}^{0} = \left| \vec{m}'\right|^2,\\
\mathscr{E}_{\mathrm{ex}}^{\textsc{d}} &= \mathcal{F}_{\alpha \beta } \left(m_\alpha  m_\beta ' - m_\alpha ' m_\beta  \right),\quad \mathscr{E}_{\mathrm{ex}}^{\textsc{a}} = \mathcal{K}_{\alpha \beta }m_\alpha m_\beta ,\\
\mathscr{E}_{\text{an}} &= - m_{\textsc{t}}^2,
\end{split}
\end{equation}
where the Einstein notation is used for summation, $K^{\text{eff}} = K+\pi M_s^2$, where the positive parameter $K$ is a magnetocrystalline anisotropy constant of easy-tangential type, the term $\pi M_s^2$ comes from the magnetostatic contribution \cite{Slastikov12, Sheka15c, Yershov15b} and $S$ is the cross-section area. 
Here and below the prime denotes the derivative with respect to the dimensionless coordinate $u  = s/\ell$ with $\ell = \sqrt{A/K^{\text{eff}}}$ being a magnetic length ($A$ is an exchange constant). The first term in the exchange energy $\mathscr{E}_{\mathrm{ex}}^{0}$ describes the common isotropic part of exchange expression which has formally the same form as for the straight wire. The second term $\mathscr{E}_{\mathrm{ex}}^{\textsc{d}}$ in the exchange energy functional is a curvature induced effective Dzyaloshinskii-Moriya interaction (DMI), which is linear with respect to curvature and torsion. The tensor of coefficients of such interaction is the dimensionless Frenet--Serret tensor \cite{Sheka15}
\begin{equation*} 
\left\|\mathcal{F}_{\alpha \beta } \right\| =
\begin{pmatrix}
0 & \varkappa & 0 \\
-\varkappa & 0 & \sigma  \\
0 &- \sigma  & 0
\end{pmatrix}.
\end{equation*}
Here $\varkappa =\kappa \ell$ and $\sigma =\tau \ell$ are the dimensionless curvature and torsion, respectively, with $\kappa$ being the curvature and $\tau$ being the torsion. The term $\mathscr{E}_{\mathrm{ex}}^{\textsc{a}}$ describes an effective anisotropy interaction, where the components of the tensor $\mathcal{K}_{\alpha \beta } = \mathcal{F}_{\alpha \nu }\mathcal{F}_{\beta \nu }$ are bilinear with respect to the curvature and the torsion,
\begin{equation*} 
\left\|\mathcal{K}_{\alpha \beta } \right\| =
\begin{pmatrix}
\varkappa ^2  & 0       & -\varkappa  \sigma  \\
0    & \varkappa ^2+\sigma ^2 & 0  \\
-\varkappa \sigma   & 0       & \sigma ^2
\end{pmatrix}.
\end{equation*}

The energy of effective anisotropy
\begin{equation*} 
\mathscr{E}_{\mathrm{eff}}^{\textsc{a}} = \mathscr{E}_{\text{an}} +  \mathscr{E}_{\mathrm{ex}}^{\textsc{a}} = \mathcal{K}_{\alpha \beta }^{\mathrm{eff}} m_{\alpha }m_{\beta }, \quad \mathcal{K}_{\alpha \beta }^{\mathrm{eff}} = \mathcal{K}_{\alpha \beta }-\delta_{\alpha,1}\delta_{\beta,1}
\end{equation*}
has a form, typical for biaxial magnets. The tensor of effective anisotropy coefficients $\mathcal{K}_{\alpha \beta }^{\mathrm{eff}}$ has non--diagonal components. This means that the homogeneous magnetization structure is not oriented along the TNB basis. One can easily diagonalize it, by using a unitary transformation (rotation in a local rectifying plane) of the vector $\vec{m}$ \eqref{eq:m-TNB}
\begin{equation*} 
\begin{split}
\vec{m} = U \vec{\widetilde{m}}, \quad &\vec{\widetilde{m}} = U^{-1}\vec{m}, \quad \vec{\widetilde{m}}=
\begin{pmatrix}
m_1, m_2, m_3
\end{pmatrix}^T
\\
& U =
\begin{pmatrix}
\cos\psi    & 0 & -\sin \psi  \\
0           & 1 & 0 \\
\sin\psi    & 0 &\cos \psi 
\end{pmatrix}.
\end{split}
\end{equation*}
By choosing the rotation angle $\psi$ as follows
\begin{equation} \label{eq:helix-psi}
\begin{split}
\psi  &= \arctan\frac{\sigma \varkappa }{\mathscr{K}_0},\\
\mathscr{K}_0 &= \frac{1+\sigma ^2-\varkappa ^2+\mathscr{K}_1}{2},\\
\mathscr{K}_1 &= \sqrt{(1-\varkappa ^2+\sigma ^2)^2+4\varkappa ^2\sigma ^2},
\end{split}
\end{equation}
one can reduce the anisotropy energy $\mathscr{E}_{\mathrm{eff}}^{\textsc{a}}$ to the form \noeqref{eq:Ean-eff}
\begin{equation}\label{eq:Ean-eff}
\begin{split}
\mathscr{E}_{\mathrm{eff}}^{\textsc{a}} &= -\mathscr{K}_1 m_1^2+\mathscr{K}_2 m_2^2,\\
\mathscr{K}_2 &= \frac{1+\varkappa ^2+\sigma ^2-\mathscr{K}_1}{2} = \frac{2\varkappa ^2}{1+\varkappa ^2+\sigma ^2+\mathscr{K}_1}.
\end{split}
\end{equation}
Here the coefficient $\mathscr{K}_1$ characterizes the strength of the effective easy-axis anisotropy  while $\mathscr{K}_2$ gives the strength of the effective easy-surface anisotropy. The direction of effective easy axis is determined by $\vec{e}_1$ and the hard axis by $\vec{e}_2$:
\begin{equation*} 
\vec{e}_1 = \vec{e}_{\textsc{t}}\cos\psi  + \vec{e}_{\textsc{b}}\sin\psi , \quad
\vec{e}_3 = -\vec{e}_{\textsc{t}}\sin\psi  + \vec{e}_{\textsc{b}}\cos\psi .
\end{equation*}
One has to note that for any finite $\psi$ the effective anisotropy direction $\vec{e}_1$ deviates from the magnetic anisotropy direction $\vec{e}_{\textsc{t}}$. Note that such a deviation vanishes for wires with zero torsion ($\sigma=0$).

Apart from effective anisotropy, the curvature and torsion show up in the effective DMI, see Eq.~\eqref{eq:energy}. In the new frame of reference ($\psi$-frame) the effective Dzyaloshinskii energy reads\footnote{In the current analysis we suppose the spatio independence of the curvature and torsion (which is adequate for the helix geometry), hence $\varkappa'=\sigma'=0$.}\noeqref{eq:ED-eff}
\begin{equation} \label{eq:ED-eff}
\begin{split}
\mathscr{E}_{\mathrm{ex}}^{\textsc{d}} &=  \mathscr{D}_1 \left( m_2 m_3'-m_3 m_2'\right) + \mathscr{D}_2 \left( m_1 m_2'-m_2 m_1'\right),\\
\mathscr{D}_1 &= 2\sigma \cos\psi  + 2\varkappa \sin\psi  = 2\sigma \frac{\mathscr{K}_0 + \varkappa^2}{\sqrt{\mathscr{K}_0^2+\sigma ^2 \varkappa ^2}},\\
\mathscr{D}_2 &= 2\varkappa \cos\psi  - 2\sigma \sin\psi  = 2\varkappa \frac{\mathscr{K}_0 - \sigma^2}{\sqrt{\mathscr{K}_0^2+\sigma ^2 \varkappa ^2}}.
\end{split}
\end{equation}
{
	\usetagform{default}
	
	Finally we get the energy in the following form of Eq.~\eqref{eq:energy-helix} of the manuscript
	\begin{equation} \label{eq:energy-helix}
	\tag{2}
	\begin{split}
	\mathscr{E} =& \underbrace{\left|\vec{m}'\right|^2}_{\text{isotropic exchange}}  \underbrace{-\mathscr{K}_1 m_1^2 + \mathscr{K}_2 m_2^2}_{\text{effective anisotropy}}\\
	& \underbrace{+ \mathscr{D}_1 \left( m_2 m_3'-m_3 m_2'\right) + \mathscr{D}_2 \left( m_1 m_2'-m_2 m_1'\right)}_{\text{effective DMI}}.
	\end{split}
	\end{equation}
}

\usetagform{suppl}

The dynamics of magnetization  is described by the Landau--Lifshitz equations for the normalized magnetization $\vec{m}$. Using the angular parametrization,
\begin{equation*} 
\vec{m} = \cos\theta~\vec{e}_1 + \sin\theta \cos\phi~\vec{e}_2 + \sin\theta \sin\phi~\vec{e}_3,
\end{equation*}
these equations can be derived from the Lagrangian
\begin{equation} \label{eq:Lagrangian}
\begin{split}
L &= K^{\text{eff}}S {\ell} \int \mathscr{L} \mathrm{d} {u}, \qquad \mathscr{L} = \mathscr{G} - \mathscr{E},\\
\mathscr{G} &= -\cos\theta \dot{\phi},\\
\mathscr{E} &= \theta '^2 + \sin^2\theta \phi '^2- \mathscr{K}_1 \cos ^2\theta+ \mathscr{K}_2 \sin ^2\theta \cos ^2\phi \\
& + \mathscr{D}_1 \sin^2\theta  \phi ' + 2\mathscr{D}_2 \sin^2\theta \cos\phi  \theta '
\end{split}
\end{equation}
and the dissipative function
\begin{equation*} 
\mathcal{F} = K^{\text{eff}}S{\ell} \int \mathscr{F} \mathrm{d} {u}, \qquad \mathscr{F} = \frac{\eta}{2} \left({\dot{\theta}}^2 + \sin^2\theta {\dot{\phi}}^2\right). 
\end{equation*}
Here and below the overdot indicates derivative with respect to the rescaled time $\bar{t}= \omega_0 t$ and $\omega_0 = \upgamma_e K^{\text{eff}}/M_s$.

\section{Static Domain Wall}

\begin{figure*}[t]
	\centering
	\includegraphics[width=\linewidth]{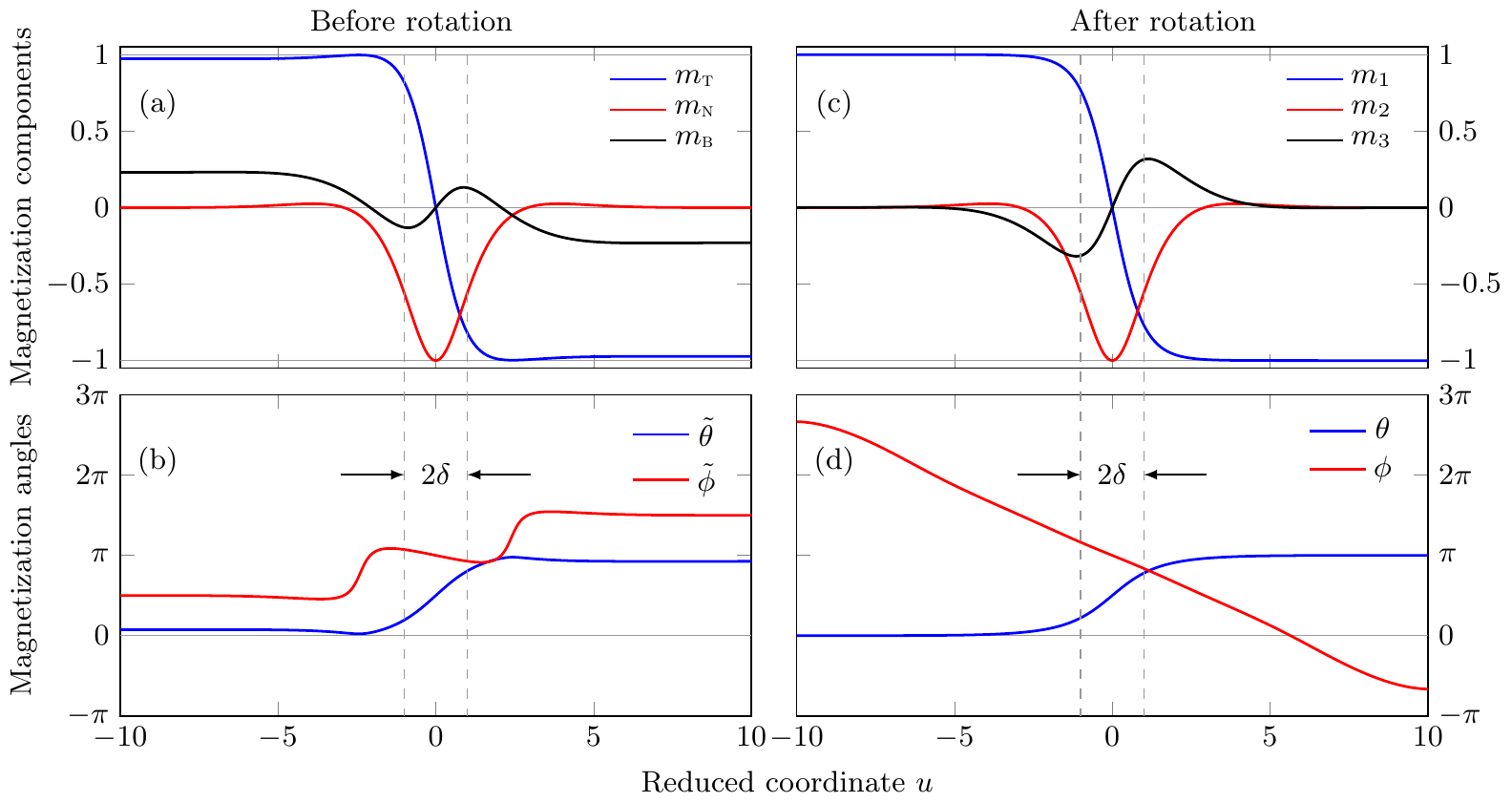}
	\caption{\textbf{Comparison of the domain wall view in the TNB and the rotated reference frame (\textsf{SLaSi} simulations for the head-to-head domain wall):}  magnetization components $m_{\textsc{t,n,b}}$ and angles $\tilde{\theta} = \arccos m_\textsc{t}$, $\tilde{\phi} = \arctan m_\textsc{b}/m_\textsc{n}$. Right column: the same in the $\psi$-frame. Parameters: $\varkappa = 0.1$, $\sigma = 0.5$, $\ell = 15a$ with $a$ being a lattice constant. Separate points are not shown due to their high density on the plots.
	}
	\label{fig:comparison}
\end{figure*}

{
	\usetagform{default}
	In the case of small enough curvature ($\varkappa \ll 1$) a static domain wall in the helix wire is well described by the expression~\eqref{eq:dw-ansatz} of the manuscript
	\begin{equation} \label{eq:dw-ansatz}
	\tag{3}
	\cos\theta ^{\text{dw}}(u) = -p\tanh \frac{u}{\delta},\qquad \phi ^{\text{dw}}(u) = {\varPhi}- \varUpsilon u,
	\end{equation}
}
where $p = \pm 1$ is a domain wall topological charge.

One can determine the magnetiochirality, \emph{i.~e.} the chirality of the magnetization structure using the Lifshitz invariant
\begin{equation} \label{eq:chirality}
\mathfrak{C} = \text{sgn }\int\limits_{-\infty}^\infty \left(m_2 m_3'-m_3 m_2'\right) \mathrm{d}u.
\end{equation}
For the domain wall \eqref{eq:dw-ansatz} one gets $\mathfrak{C} = - \text{sgn }\varUpsilon$.

Let us compare the magnetization distribution in $\psi$-frame [Eq.~\eqref{eq:dw-ansatz}] with the magnetization distribution in the TNB reference frame which are connected by the following relations:
\begin{equation*} 
\begin{split}
m_\textsc{t} &= m_1\cos\psi - m_3 \sin\psi, \\
m_\textsc{n} &= m_2, \\
m_\textsc{b} &= m_1\sin\psi + m_3\cos\psi ,
\end{split}
\end{equation*}
or, in the angular parametrization,
\begin{equation*}
\begin{split}
\cos\tilde{\theta} & = m_\textsc{t} = \cos\theta \cos\psi - \sin\theta \sin\phi \sin \psi,\\
\tan \tilde{\phi} & = \dfrac{m_\textsc{b}}{m_\textsc{n}} = \dfrac{\cos\theta \sin\psi + \sin\theta \sin\phi \cos \psi}{\sin\theta \cos\phi}.
\end{split}
\end{equation*}

Comparison of the domain wall shapes in two above mentioned reference frames (magnetization components and angles) is shown in Fig.~\ref{fig:comparison}, obtained from \textsf{SLaSi} simulations\cite{slasi}, c.\,f. Fig.~4 of the manuscript, see Methods for details. Figures~\ref{fig:comparison}(a) and (b) clearly pronounce that the ground state is never strictly tangential one: the component $m_\textsc{b}$ and, therefore, the angle $\tilde{\phi}$ are nonzero far from the domain wall. In the left and the right domains the magnetization states are $\tilde{\theta} = \psi$, $\tilde{\phi} = \pi/2\mod 2\pi$ and $\tilde{\theta} = \pi - \psi$, $\tilde{\phi} = 3\pi/2\mod 2\pi$ respectively. Inside the domain wall a bend of the the $\tilde{\phi}(u)$ profile appears. In the rotated reference frame domain wall structure significantly simplifies: $\phi(u)$ has a shape close to linear function and $m_2$, $m_3$ components becomes localized. 

\begin{figure*}[t]
	\centering
	\includegraphics[width=\linewidth]{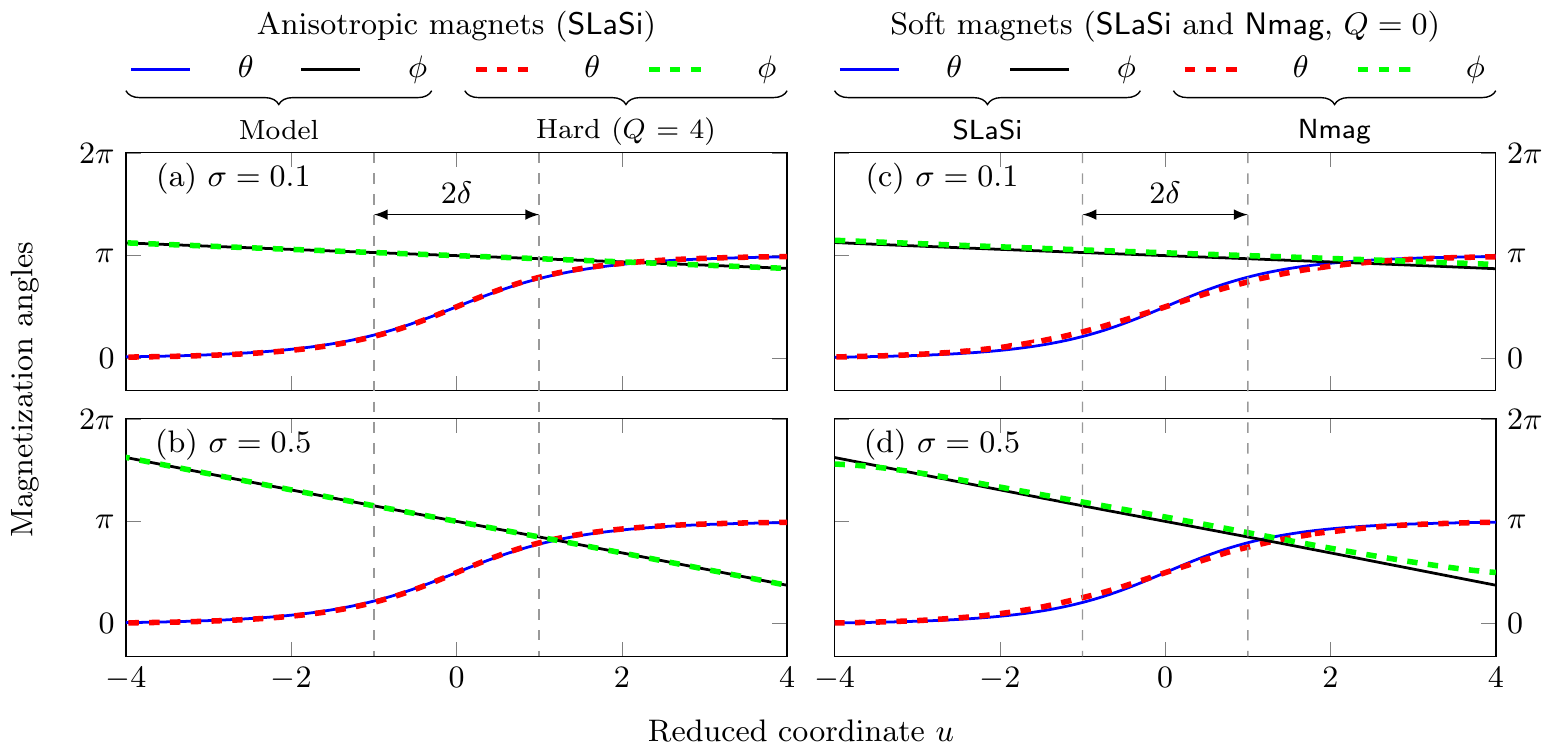}
	\caption{\textbf{Influence of magnetostatics on the static domain wall:} Magnetization angles in the $\psi$-frame for the head-to-head domain wall for $\varkappa=0.1$ and different $\sigma$. Simulations without magnetostatics (model, solid lines) and of magnetically hard magnets ($Q = 4$, dashed lines) for $\sigma = 0.1$ (a) and $\sigma = 0.5$ (b), spin-lattice simulations in \textsf{SLaSi}. Simulations of magnetically soft magnets ($Q = 0$), spin-lattice simulations in \textsf{SLaSi} (solid lines) and micromagnetic simulations in \textsf{Nmag} (dashed lines) for $\sigma = 0.1$ (a) and $\sigma = 0.5$ (b). Magnetic parameters correspond to the magnetic length~$\ell = 15a$. Rotation angle $\psi_\text{sim}$ is determined from simulations for all curves where magnetostatics is taken into account [$|\psi - \psi_\text{sim}| < 0.004$, where $\psi$ is determined by Eq.~\eqref{eq:helix-psi}].}
	\label{fig:comparisonDip}
\end{figure*}

Figure~\ref{fig:comparisonDip} shows a comparison of domain wall structure for different values of quality factor $Q = K/2\pi M_\textsc{s}^2$: $Q = 0$ and $Q = 4$ in spin-lattice simulations with micromagnetic simulations and model where dipolar interaction is replaced by easy--tangential anisotropy only, see Methods for details.

\section{Effective equations of the domain wall motion under the influence of Rashba torque}

\usetagform{default}

In order to derive effective equations of the domain wall motion we use {generalized collective coordinate $q$--$\varPhi$ approach \cite{Kravchuk14} based on the effective Lagrangian formalism. We start from the travelling wave Ansatz (see Eq.~\eqref{eq:dw-ansatz-dynamics} of the manuscript):
	\begin{equation} \label{eq:dw-ansatz-dynamics}
	\tag{6}
	\begin{split}
	\cos\theta ^{\text{dw}}(u,\bar{t}) &= -p\tanh \frac{u-q(\bar{t})}{\delta},\\
	\phi ^{\text{dw}}(u,\bar{t})   &= \varPhi(\bar{t})-\varUpsilon \left[u-q(\bar{t})\right].
	\end{split}
	\end{equation}
	
	\usetagform{suppl}
	
	One can derive the effective Lagrangian of the system by inserting this Ansatz into the full Lagrangian \eqref{eq:Lagrangian}, and calculating the integral over the dimensionless coordinate $u$. Then the effective Lagrangian, normalized by $K S \ell$ reads $L^{\text{eff}} = G^{\text{eff}}-E^{\text{eff}}$ with effective gyroscopical term $G^{\text{eff}} = 2p\varPhi\dot{q}$ and the effective energy, cf. Eq.~(M1) of the manuscript:
	\begin{equation*} 
	\begin{split}
	E^{\text{eff}} &=\dfrac{2}{\delta}  +\delta\left[2\mathscr{K}_1  + 2 \varUpsilon^2 + \mathscr{K}_2 \left(1+\mathscr{C}_1 \cos2\varPhi\right)\right]\\
	& - 2\delta \mathscr{D}_1 \varUpsilon  + p \mathscr{C}_2 \mathscr{D}_2 \cos\varPhi - 4 p h q \sin\psi,\\
	\mathscr{C}_1&=  \frac{\pi \delta \varUpsilon}{\sinh (\pi \delta \varUpsilon)}, \qquad \mathscr{C}_2 = \frac{\pi   (1 + \delta^2 \varUpsilon^2)}{ \cosh (\pi \delta \varUpsilon/2)}.
	\end{split}
	\end{equation*}
	In the same way one can derive an effective dissipative function
	\begin{equation*} 
	F^{\text{eff}} = \eta \left[\frac{{\dot q}^2}{\delta} + \delta \left(\dot{\varPhi} + \varUpsilon \dot{q}\right)^2\right].
	\end{equation*}
	From the Euler-Lagrange-Rayleigh equations (M2) for the set of variables $X_i = \left\{q,\varPhi \right\}$ we obtain finally
	\begin{equation} \label{eq:ELR}
	\begin{split}
	&\dot{\varPhi} \left(p+\eta \delta \varUpsilon \right) + \frac{\eta}{\delta} \dot{q} \left(1+\delta^2 \varUpsilon^2\right) = 2ph\sin\psi,\\
	& \eta \delta \dot{\varPhi} - \dot{q} \left(p-\eta \delta \varUpsilon \right) = \mathscr{C}_1 \mathscr{K}_2 \delta \sin 2\varPhi + \frac{p}{2} \mathscr{C}_2 \mathscr{D}_2 \sin \varPhi.
	\end{split}
	\end{equation}
	
	\renewcommand{\figurename}{Video}
	\setcounter{figure}{0}
%

\begin{figure}[t]
\begin{center}
	\includemedia[
	label=steady,
	activate=onclick,
	width=0.9\columnwidth,height=0.5125\columnwidth,
	flashvars={modestbranding=1 
		& autohide=1 
		& showinfo=0 
		& rel=0 
	}
	]{\includegraphics[width=\columnwidth]{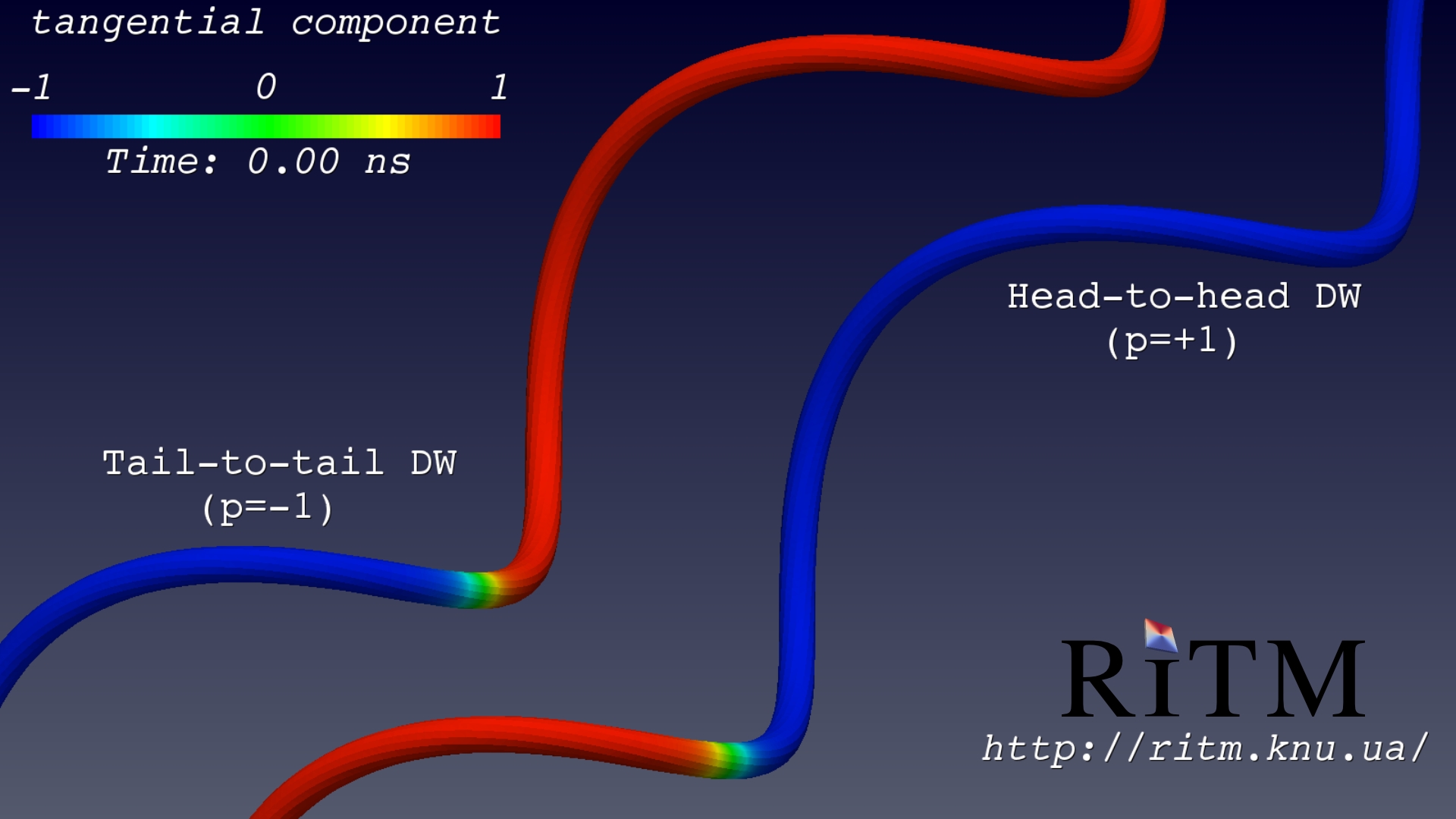}}
	{https://www.youtube.com/v/nS2b0E5_AT8?rel=0}
\end{center}
	\caption{Motion of the head-to-head ($p=+1$) and tail-to-tail ($p = -1$) domain walls in helices with curvature $\varkappa = 0.1$ and torsion $\sigma = 0.1$ under the action of the Rashba field $h=0.02$.}
	\label{mov:motion}
\end{figure}

	The effective equations of motion \eqref{eq:ELR} provide the domain wall motion with the finite velocity
	\usetagform{default}
	(see Eq.~\eqref{eq:velocity} of the manuscript):
	\begin{equation} \label{eq:velocity}
	\tag{7}
	v \equiv \frac{\mathrm{d}q}{\mathrm{d}\bar{t}} (\bar{t}\to \infty) = \frac{2 p h \delta }{\eta} \cdot \dfrac{\sin\psi}{1+\delta^2\varUpsilon^2}.
	\end{equation}
	\usetagform{suppl}%
	The motion of domain walls in helices with different chiralities is illustrated by Supplementary Video~\ref{mov:motion}.
	
	The stationary phase $\varPhi =\text{const}$ can be found from the equation:
	\begin{equation*} 
	2\mathscr{C}_1 \mathscr{K}_2 \delta \sin 2\varPhi + p \mathscr{C}_2 \mathscr{D}_2 \sin \varPhi = - \frac{4 p h \delta \sin\psi}{\eta(1+\delta^2 \varUpsilon^2)} \left(p-\eta \delta \varUpsilon \right).
	\end{equation*}
	In the case $\varkappa, |\sigma| \ll1$, one gets
	\begin{equation*} 
	\varPhi \approx \varPhi_0 + \frac{2h \sigma}{\pi\eta} .
	\end{equation*}


\begin{thebibliography}{10}
	\expandafter\ifx\csname url\endcsname\relax
	\def\url#1{\texttt{#1}}\fi
	\expandafter\ifx\csname urlprefix\endcsname\relax\def\urlprefix{URL }\fi
	\providecommand{\bibinfo}[2]{#2}
	\providecommand{\eprint}[2][]{\url{#2}}
	
	\bibitem{Albrecht05}
	\bibinfo{author}{Albrecht, M.} \emph{et~al.}
	\newblock \bibinfo{title}{Magnetic multilayers on nanospheres}.
	\newblock \emph{\bibinfo{journal}{Nat Mater}} \textbf{\bibinfo{volume}{4}},
	\bibinfo{pages}{203--206} (\bibinfo{year}{2005}).
	\newblock \urlprefix\url{http://dx.doi.org/10.1038/nmat1324}.
	
	\bibitem{Ulbrich06}
	\bibinfo{author}{Ulbrich, T.~C.} \emph{et~al.}
	\newblock \bibinfo{title}{Magnetization reversal in a novel gradient
		nanomaterial}.
	\newblock \emph{\bibinfo{journal}{Phys. Rev. Lett.}}
	\textbf{\bibinfo{volume}{96}}, \bibinfo{pages}{077202}
	(\bibinfo{year}{2006}).
	\newblock
	\urlprefix\url{http://link.aps.org/doi/10.1103/PhysRevLett.96.077202}.
	
	\bibitem{Hertel13a}
	\bibinfo{author}{Hertel, R.}
	\newblock \bibinfo{title}{Curvature--induced magnetochirality}.
	\newblock \emph{\bibinfo{journal}{SPIN}} \textbf{\bibinfo{volume}{03}},
	\bibinfo{pages}{1340009} (\bibinfo{year}{2013}).
	\newblock \urlprefix\url{http://dx.doi.org/10.1142/S2010324713400092}.
	
	\bibitem{Streubel14a}
	\bibinfo{author}{Streubel, R.} \emph{et~al.}
	\newblock \bibinfo{title}{Imaging of buried {3D} magnetic rolled-up
		nanomembranes}.
	\newblock \emph{\bibinfo{journal}{Nano Lett.}} \textbf{\bibinfo{volume}{14}},
	\bibinfo{pages}{3981--3986} (\bibinfo{year}{2014}).
	\newblock \urlprefix\url{http://dx.doi.org/10.1021/nl501333h}.
	
	\bibitem{Streubel14}
	\bibinfo{author}{Streubel, R.} \emph{et~al.}
	\newblock \bibinfo{title}{Magnetic microstructure of rolled-up single-layer
		ferromagnetic nanomembranes}.
	\newblock \emph{\bibinfo{journal}{Advanced Materials}}
	\textbf{\bibinfo{volume}{26}}, \bibinfo{pages}{316--323}
	(\bibinfo{year}{2014}).
	\newblock \urlprefix\url{http://dx.doi.org/10.1002/adma.201303003}.
	
	\bibitem{Streubel15a}
	\bibinfo{author}{Streubel, R.} \emph{et~al.}
	\newblock \bibinfo{title}{Retrieving spin textures on curved magnetic thin
		films with full-field soft {X}-ray microscopies}.
	\newblock \emph{\bibinfo{journal}{Nat Comms}} \textbf{\bibinfo{volume}{6}},
	\bibinfo{pages}{7612} (\bibinfo{year}{2015}).
	\newblock \urlprefix\url{http://dx.doi.org/10.1038/ncomms8612}.
	
	\bibitem{Nielsch01}
	\bibinfo{author}{Nielsch, K.} \emph{et~al.}
	\newblock \bibinfo{title}{Hexagonally ordered 100~{nm} period nickel nanowire
		arrays}.
	\newblock \emph{\bibinfo{journal}{Applied Physics Letters}}
	\textbf{\bibinfo{volume}{79}}, \bibinfo{pages}{1360} (\bibinfo{year}{2001}).
	\newblock \urlprefix\url{http://dx.doi.org/10.1063/1.1399006}.
	
	\bibitem{Buchter13a}
	\bibinfo{author}{Buchter, A.}, \bibinfo{author}{Nagel, J.} \&
	\bibinfo{author}{R\"uffer}.
	\newblock \bibinfo{title}{Reversal mechanism of an individual ni nanotube
		simultaneously studied by torque and squid magnetometry}.
	\newblock \emph{\bibinfo{journal}{Phys. Rev. Lett.}}
	\textbf{\bibinfo{volume}{111}}, \bibinfo{pages}{067202}
	(\bibinfo{year}{2013}).
	\newblock
	\urlprefix\url{http://link.aps.org/doi/10.1103/PhysRevLett.111.067202}.
	
	\bibitem{Rueffer12}
	\bibinfo{author}{R\"uffer, D.} \emph{et~al.}
	\newblock \bibinfo{title}{Magnetic states of an individual ni nanotube probed
		by anisotropic magnetoresistance}.
	\newblock \emph{\bibinfo{journal}{Nanoscale}} \textbf{\bibinfo{volume}{4}},
	\bibinfo{pages}{4989} (\bibinfo{year}{2012}).
	\newblock \urlprefix\url{http://dx.doi.org/10.1039/C2NR31086D}.
	
	\bibitem{Weber12}
	\bibinfo{author}{Weber, D.~P.} \emph{et~al.}
	\newblock \bibinfo{title}{Cantilever magnetometry of individual {Ni}
		nanotubes}.
	\newblock \emph{\bibinfo{journal}{Nano Letters}} \textbf{\bibinfo{volume}{12}},
	\bibinfo{pages}{6139--6144} (\bibinfo{year}{2012}).
	\newblock \urlprefix\url{http://pubs.acs.org/doi/abs/10.1021/nl302950u}.
	\newblock \eprint{http://pubs.acs.org/doi/pdf/10.1021/nl302950u}.
	
	\bibitem{Dietrich08}
	\bibinfo{author}{Dietrich, C.} \emph{et~al.}
	\newblock \bibinfo{title}{Influence of perpendicular magnetic fields on the
		domain structure of permalloy microstructures grown on thin membranes}.
	\newblock \emph{\bibinfo{journal}{Phys. Rev. B}} \textbf{\bibinfo{volume}{77}},
	\bibinfo{pages}{174427} (\bibinfo{year}{2008}).
	\newblock \urlprefix\url{http://link.aps.org/doi/10.1103/PhysRevB.77.174427}.
	
	\bibitem{Otalora12a}
	\bibinfo{author}{Ot\'{a}lora, J.}, \bibinfo{author}{L\'{o}pez-L\'{o}pez, J.},
	\bibinfo{author}{Vargas, P.} \& \bibinfo{author}{Landeros, P.}
	\newblock \bibinfo{title}{Chirality switching and propagation control of a
		vortex domain wall in ferromagnetic nanotubes}.
	\newblock \emph{\bibinfo{journal}{Applied Physics Letters}}
	\textbf{\bibinfo{volume}{100}}, \bibinfo{pages}{072407}
	(\bibinfo{year}{2012}).
	\newblock
	\urlprefix\url{http://scitation.aip.org/content/aip/journal/apl/100/7/10.1063/1.3687154}.
	
	\bibitem{Kravchuk12a}
	\bibinfo{author}{Kravchuk, V.~P.} \emph{et~al.}
	\newblock \bibinfo{title}{Out-of-surface vortices in spherical shells}.
	\newblock \emph{\bibinfo{journal}{Phys. Rev. B}} \textbf{\bibinfo{volume}{85}},
	\bibinfo{pages}{144433} (\bibinfo{year}{2012}).
	\newblock \urlprefix\url{http://link.aps.org/doi/10.1103/PhysRevB.85.144433}.
	
	\bibitem{Smith11}
	\bibinfo{author}{Smith, E.~J.}, \bibinfo{author}{Makarov, D.},
	\bibinfo{author}{Sanchez, S.}, \bibinfo{author}{Fomin, V.~M.} \&
	\bibinfo{author}{Schmidt, O.~G.}
	\newblock \bibinfo{title}{Magnetic microhelix coil structures}.
	\newblock \emph{\bibinfo{journal}{Phys. Rev. Lett.}}
	\textbf{\bibinfo{volume}{107}}, \bibinfo{pages}{097204}
	(\bibinfo{year}{2011}).
	\newblock
	\urlprefix\url{http://link.aps.org/doi/10.1103/PhysRevLett.107.097204}.
	
	\bibitem{Pylypovskyi15b}
	\bibinfo{author}{Pylypovskyi, O.~V.} \emph{et~al.}
	\newblock \bibinfo{title}{Coupling of chiralities in spin and physical spaces:
		{T}he {M}\"obius ring as a case study}.
	\newblock \emph{\bibinfo{journal}{Phys. Rev. Lett.}}
	\textbf{\bibinfo{volume}{114}}, \bibinfo{pages}{197204}
	(\bibinfo{year}{2015}).
	\newblock
	\urlprefix\url{http://link.aps.org/doi/10.1103/PhysRevLett.114.197204}.
	
	\bibitem{Gaididei14}
	\bibinfo{author}{Gaididei, Y.}, \bibinfo{author}{Kravchuk, V.~P.} \&
	\bibinfo{author}{Sheka, D.~D.}
	\newblock \bibinfo{title}{Curvature effects in thin magnetic shells}.
	\newblock \emph{\bibinfo{journal}{Phys. Rev. Lett.}}
	\textbf{\bibinfo{volume}{112}}, \bibinfo{pages}{257203}
	(\bibinfo{year}{2014}).
	\newblock
	\urlprefix\url{http://link.aps.org/doi/10.1103/PhysRevLett.112.257203}.
	
	\bibitem{Sheka15}
	\bibinfo{author}{Sheka, D.~D.}, \bibinfo{author}{Kravchuk, V.~P.} \&
	\bibinfo{author}{Gaididei, Y.}
	\newblock \bibinfo{title}{Curvature effects in statics and dynamics of low
		dimensional magnets}.
	\newblock \emph{\bibinfo{journal}{Journal of Physics A: Mathematical and
			Theoretical}} \textbf{\bibinfo{volume}{48}}, \bibinfo{pages}{125202}
	(\bibinfo{year}{2015}).
	\newblock \urlprefix\url{http://stacks.iop.org/1751-8121/48/i=12/a=125202}.
	
	\bibitem{Moench11}
	\bibinfo{author}{M{\"o}nch, I.} \emph{et~al.}
	\newblock \bibinfo{title}{Rolled-up magnetic sensor: Nanomembrane architecture
		for in-flow detection of magnetic objects}.
	\newblock \emph{\bibinfo{journal}{ACS Nano}} \textbf{\bibinfo{volume}{5}},
	\bibinfo{pages}{7436--7442} (\bibinfo{year}{2011}).
	\newblock \urlprefix\url{http://dx.doi.org/10.1021/nn202351j}.
	
	\bibitem{Muller12}
	\bibinfo{author}{M\"uller, C.} \emph{et~al.}
	\newblock \bibinfo{title}{Towards compact three-dimensional
		magnetoelectronics--magnetoresistance in rolled-up {Co/Cu} nanomembranes}.
	\newblock \emph{\bibinfo{journal}{Applied Physics Letters}}
	\textbf{\bibinfo{volume}{100}}, \bibinfo{pages}{022409}
	(\bibinfo{year}{2012}).
	\newblock \urlprefix\url{http://dx.doi.org/10.1063/1.3676269}.
	
	\bibitem{Balhorn10}
	\bibinfo{author}{Balhorn, F.} \emph{et~al.}
	\newblock \bibinfo{title}{Spin-wave interference in three-dimensional rolled-up
		ferromagnetic microtubes}.
	\newblock \emph{\bibinfo{journal}{Physical Review Letters}}
	\textbf{\bibinfo{volume}{104}}, \bibinfo{pages}{037205}
	(\bibinfo{year}{2010}).
	\newblock \urlprefix\url{http://dx.doi.org/10.1103/PhysRevLett.104.037205}.
	
	\bibitem{Balhorn12}
	\bibinfo{author}{Balhorn, F.}, \bibinfo{author}{Jeni, S.},
	\bibinfo{author}{Hansen, W.}, \bibinfo{author}{Heitmann, D.} \&
	\bibinfo{author}{Mendach, S.}
	\newblock \bibinfo{title}{Axial and azimuthal spin-wave eigenmodes in rolled-up
		permalloy stripes}.
	\newblock \emph{\bibinfo{journal}{Applied Physics Letters}}
	\textbf{\bibinfo{volume}{100}}, \bibinfo{pages}{222402}
	(\bibinfo{year}{2012}).
	\newblock \urlprefix\url{http://dx.doi.org/10.1063/1.3700809}.
	
	\bibitem{Liu99a}
	\bibinfo{author}{Liu, L.}, \bibinfo{author}{Ioannides, A.} \&
	\bibinfo{author}{Streit, M.}
	\newblock \bibinfo{title}{Single trial analysis of neurophysiological
		correlates of the recognition of complex objects and facial expressions of
		emotion.}
	\newblock \emph{\bibinfo{journal}{Brain Topography}}
	\textbf{\bibinfo{volume}{11}}, \bibinfo{pages}{291--303}
	(\bibinfo{year}{1999}).
	
	\bibitem{Dumas13}
	\bibinfo{author}{Dumas, T.} \emph{et~al.}
	\newblock \bibinfo{title}{Meg evidence for dynamic amygdala modulations by gaze
		and facial emotions}.
	\newblock \emph{\bibinfo{journal}{PLoS ONE}} \textbf{\bibinfo{volume}{8}},
	\bibinfo{pages}{e74145} (\bibinfo{year}{2013}).
	\newblock \urlprefix\url{http://dx.doi.org/10.1371%2Fjournal.pone.0074145}.
		
		\bibitem{Karnaushenko15a}
		\bibinfo{author}{Karnaushenko, D.} \emph{et~al.}
		\newblock \bibinfo{title}{Self-assembled on-chip-integrated giant
			magneto-impedance sensorics}.
		\newblock \emph{\bibinfo{journal}{Adv. Mater.}} \bibinfo{pages}{n/a--n/a}
		(\bibinfo{year}{2015}).
		\newblock \urlprefix\url{http://dx.doi.org/10.1002/adma.201503127}.
		
		\bibitem{Parkin08}
		\bibinfo{author}{Parkin, S. S.~P.}, \bibinfo{author}{Hayashi, M.} \&
		\bibinfo{author}{Thomas, L.}
		\newblock \bibinfo{title}{Magnetic domain-wall racetrack memory}.
		\newblock \emph{\bibinfo{journal}{Science}} \textbf{\bibinfo{volume}{320}},
		\bibinfo{pages}{190--194} (\bibinfo{year}{2008}).
		
		\bibitem{Yan10}
		\bibinfo{author}{Yan, M.}, \bibinfo{author}{K\'akay, A.},
		\bibinfo{author}{Gliga, S.} \& \bibinfo{author}{Hertel, R.}
		\newblock \bibinfo{title}{Beating the walker limit with massless domain walls
			in cylindrical nanowires}.
		\newblock \emph{\bibinfo{journal}{Phys. Rev. Lett.}}
		\textbf{\bibinfo{volume}{104}}, \bibinfo{pages}{057201}
		(\bibinfo{year}{2010}).
		\newblock
		\urlprefix\url{http://link.aps.org/doi/10.1103/PhysRevLett.104.057201}.
		
		\bibitem{Catalan12}
		\bibinfo{author}{Catalan, G.}, \bibinfo{author}{Seidel, J.},
		\bibinfo{author}{Ramesh, R.} \& \bibinfo{author}{Scott, J.~F.}
		\newblock \bibinfo{title}{Domain wall nanoelectronics}.
		\newblock \emph{\bibinfo{journal}{Rev. Mod. Phys.}}
		\textbf{\bibinfo{volume}{84}}, \bibinfo{pages}{119--156}
		(\bibinfo{year}{2012}).
		\newblock \urlprefix\url{http://dx.doi.org/10.1103/RevModPhys.84.119}.
		
		\bibitem{Hayashi07}
		\bibinfo{author}{Hayashi, M.}, \bibinfo{author}{Thomas, L.},
		\bibinfo{author}{Rettner, C.}, \bibinfo{author}{Moriya, R.} \&
		\bibinfo{author}{Parkin, S. S.~P.}
		\newblock \bibinfo{title}{Direct observation of the coherent precession of
			magnetic domain walls propagating along permalloy nanowires}.
		\newblock \emph{\bibinfo{journal}{Nat Phys}} \textbf{\bibinfo{volume}{3}},
		\bibinfo{pages}{21--25} (\bibinfo{year}{2007}).
		\newblock \urlprefix\url{http://dx.doi.org/10.1038/nphys464}.
		
		\bibitem{Allwood02}
		\bibinfo{author}{Allwood, D.~A.} \emph{et~al.}
		\newblock \bibinfo{title}{Submicrometer ferromagnetic {NOT} gate and shift
			register}.
		\newblock \emph{\bibinfo{journal}{Science}} \textbf{\bibinfo{volume}{296}},
		\bibinfo{pages}{2003--2006} (\bibinfo{year}{2002}).
		\newblock \urlprefix\url{http://dx.doi.org/10.1126/science.1070595}.
		
		\bibitem{Allwood05}
		\bibinfo{author}{Allwood, D.~A.} \emph{et~al.}
		\newblock \bibinfo{title}{Magnetic domain--wall logic}.
		\newblock \emph{\bibinfo{journal}{Science}} \textbf{\bibinfo{volume}{309}},
		\bibinfo{pages}{1688--1692} (\bibinfo{year}{2005}).
		\newblock \urlprefix\url{http://dx.doi.org/10.1126/science.1108813}.
		
		\bibitem{Vazquez15}
		\bibinfo{author}{V{\'a}zquez, M.}
		\newblock \emph{\bibinfo{title}{Magnetic nano- and microwires: design,
				synthesis, properties and applications}} (\bibinfo{publisher}{Woodhead
			Publishing is an imprint of Elsevier}, \bibinfo{address}{Cambridge, UK},
		\bibinfo{year}{2015}).
		
		\bibitem{Manchon14}
		\bibinfo{author}{Manchon, A.}
		\newblock \bibinfo{title}{Spin--orbitronics: {A} new moment for {B}erry}.
		\newblock \emph{\bibinfo{journal}{Nature Physics}}
		\textbf{\bibinfo{volume}{10}}, \bibinfo{pages}{340--341}
		(\bibinfo{year}{2014}).
		\newblock \urlprefix\url{http://dx.doi.org/10.1038/nphys2957}.
		
		\bibitem{Kuschel15}
		\bibinfo{author}{Kuschel, T.} \& \bibinfo{author}{Reiss, G.}
		\newblock \bibinfo{title}{Spin orbitronics: Charges ride the spin wave}.
		\newblock \emph{\bibinfo{journal}{Nature Nanotechnology}}
		\textbf{\bibinfo{volume}{10}}, \bibinfo{pages}{22--24}
		(\bibinfo{year}{2015}).
		\newblock \urlprefix\url{http://dx.doi.org/10.1038/nnano.2014.279}.
		
		\bibitem{Miron10}
		\bibinfo{author}{Miron, I.~M.} \emph{et~al.}
		\newblock \bibinfo{title}{Current-driven spin torque induced by the {R}ashba
			effect in a ferromagnetic metal layer}.
		\newblock \emph{\bibinfo{journal}{Nat Mater}}  (\bibinfo{year}{2010}).
		\newblock \urlprefix\url{http://dx.doi.org/10.1038/nmat2613}.
		
		\bibitem{Martinez13}
		\bibinfo{author}{Martinez, E.}, \bibinfo{author}{Emori, S.} \&
		\bibinfo{author}{Beach, G. S.~D.}
		\newblock \bibinfo{title}{{Current-driven domain wall motion along high
				perpendicular anisotropy multilayers: The role of the Rashba field, the spin
				Hall effect, and the Dzyaloshinskii-Moriya interaction}}.
		\newblock \emph{\bibinfo{journal}{Applied Physics Letters}}
		\textbf{\bibinfo{volume}{103}}, \bibinfo{pages}{072406}
		(\bibinfo{year}{2013}).
		\newblock \urlprefix\url{http://dx.doi.org/10.1063/1.4818723}.
		
		\bibitem{Khvalkovskiy13}
		\bibinfo{author}{Khvalkovskiy, A.~V.} \emph{et~al.}
		\newblock \bibinfo{title}{Matching domain-wall configuration and spin-orbit
			torques for efficient domain-wall motion}.
		\newblock \emph{\bibinfo{journal}{Phys. Rev. B}} \textbf{\bibinfo{volume}{87}},
		\bibinfo{pages}{020402} (\bibinfo{year}{2013}).
		\newblock \urlprefix\url{http://link.aps.org/doi/10.1103/PhysRevB.87.020402}.
		
		\bibitem{Sheka15c}
		\bibinfo{author}{Sheka, D.~D.}, \bibinfo{author}{Kravchuk, V.~P.},
		\bibinfo{author}{Yershov, K.~V.} \& \bibinfo{author}{Gaididei, Y.}
		\newblock \bibinfo{title}{Torsion-induced effects in magnetic nanowires}.
		\newblock \emph{\bibinfo{journal}{Phys. Rev. B}} \textbf{\bibinfo{volume}{92}},
		\bibinfo{pages}{054417} (\bibinfo{year}{2015}).
		\newblock \urlprefix\url{http://link.aps.org/doi/10.1103/PhysRevB.92.054417}.
		
		\bibitem{Slastikov12}
		\bibinfo{author}{Slastikov, V.~V.} \& \bibinfo{author}{Sonnenberg, C.}
		\newblock \bibinfo{title}{Reduced models for ferromagnetic nanowires}.
		\newblock \emph{\bibinfo{journal}{IMA Journal of Applied Mathematics}}
		\textbf{\bibinfo{volume}{77}}, \bibinfo{pages}{220--235}
		(\bibinfo{year}{2012}).
		\newblock \urlprefix\url{http://dx.doi.org/10.1093/imamat/hxr019}.
		
		\bibitem{Yershov15b}
		\bibinfo{author}{Yershov, K.~V.}, \bibinfo{author}{Kravchuk, V.~P.},
		\bibinfo{author}{Sheka, D.~D.} \& \bibinfo{author}{Gaididei, Y.}
		\newblock \bibinfo{title}{Curvature-induced domain wall pinning}.
		\newblock \emph{\bibinfo{journal}{Phys. Rev. B}} \textbf{\bibinfo{volume}{92}},
		\bibinfo{pages}{104412} (\bibinfo{year}{2015}).
		\newblock \urlprefix\url{http://link.aps.org/doi/10.1103/PhysRevB.92.104412}.
		
		\bibitem{Dzyaloshinsky58}
		\bibinfo{author}{Dzyaloshinsky, I.}
		\newblock \bibinfo{title}{A thermodynamic theory of ``weak'' ferromagnetism of
			antiferromagnetics}.
		\newblock \emph{\bibinfo{journal}{Journal of Physics and Chemistry of Solids}}
		\textbf{\bibinfo{volume}{4}}, \bibinfo{pages}{241 -- 255}
		(\bibinfo{year}{1958}).
		\newblock
		\urlprefix\url{http://www.sciencedirect.com/science/article/pii/0022369758900763}.
		
		\bibitem{Moriya60}
		\bibinfo{author}{Moriya, T.}
		\newblock \bibinfo{title}{Anisotropic superexchange interaction and weak
			ferromagnetism}.
		\newblock \emph{\bibinfo{journal}{Phys. Rev.}} \textbf{\bibinfo{volume}{120}},
		\bibinfo{pages}{91--98} (\bibinfo{year}{1960}).
		\newblock \urlprefix\url{http://link.aps.org/abstract/PR/v120/p91}.
		
		\bibitem{Crepieux98}
		\bibinfo{author}{Cr{'e}pieux, A.} \& \bibinfo{author}{Lacroix, C.}
		\newblock \bibinfo{title}{{Dzyaloshinsky--Moriya interactions induced by
				symmetry breaking at a surface}}.
		\newblock \emph{\bibinfo{journal}{Journal of Magnetism and Magnetic Materials}}
		\textbf{\bibinfo{volume}{182}}, \bibinfo{pages}{341--349}
		(\bibinfo{year}{1998}).
		\newblock \urlprefix\url{http://dx.doi.org/10.1016/S0304-8853(97)01044-5}.
		
		\bibitem{Kravchuk14}
		\bibinfo{author}{Kravchuk, V.~P.}
		\newblock \bibinfo{title}{Influence of {D}zialoshinskii--{M}oriya interaction
			on static and dynamic properties of a transverse domain wall}.
		\newblock \emph{\bibinfo{journal}{Journal of Magnetism and Magnetic Materials}}
		\textbf{\bibinfo{volume}{367}}, \bibinfo{pages}{9} (\bibinfo{year}{2014}).
		\newblock \urlprefix\url{http://dx.doi.org/10.1016/j.jmmm.2014.04.073}.
		
		\bibitem{slasi}
		\bibinfo{title}{\texttt{SLaSi} spin--lattice simulations package}.
		\newblock \urlprefix\url{http://slasi.rpd.univ.kiev.ua}.
		
		\bibitem{Fischbacher07}
		\bibinfo{author}{Fischbacher, T.}, \bibinfo{author}{Franchin, M.},
		\bibinfo{author}{Bordignon, G.} \& \bibinfo{author}{Fangohr, H.}
		\newblock \bibinfo{title}{A systematic approach to multiphysics extensions of
			finite-element-based micromagnetic simulations: Nmag}.
		\newblock \emph{\bibinfo{journal}{IEEE Trans. Magn.}}
		\textbf{\bibinfo{volume}{43}}, \bibinfo{pages}{2896--2898}
		(\bibinfo{year}{2007}).
		\newblock \urlprefix\url{http://dx.doi.org/10.1109/TMAG.2007.893843}.
		
		\bibitem{Obata08}
		\bibinfo{author}{Obata, K.} \& \bibinfo{author}{Tatara, G.}
		\newblock \bibinfo{title}{{Current-induced domain wall motion in Rashba
				spin-orbit system}}.
		\newblock \emph{\bibinfo{journal}{Phys. Rev. B}} \textbf{\bibinfo{volume}{77}},
		\bibinfo{pages}{214429} (\bibinfo{year}{2008}).
		\newblock \urlprefix\url{http://link.aps.org/abstract/PRB/v77/e214429}.
		
		\bibitem{Manchon09}
		\bibinfo{author}{Manchon, A.} \& \bibinfo{author}{Zhang, S.}
		\newblock \bibinfo{title}{Theory of spin torque due to spin-orbit coupling}.
		\newblock \emph{\bibinfo{journal}{Phys. Rev. B}} \textbf{\bibinfo{volume}{79}},
		\bibinfo{pages}{094422} (\bibinfo{year}{2009}).
		\newblock \urlprefix\url{http://dx.doi.org/10.1103/PhysRevB.79.094422}.
		
		\bibitem{Sobolev95}
		\bibinfo{author}{Sobolev, V.}, \bibinfo{author}{Huang, H.} \&
		\bibinfo{author}{Chen, S.}
		\newblock \bibinfo{title}{Domain wall dynamics in the presence of an external
			magnetic field normal to the anisotropy axis}.
		\newblock \emph{\bibinfo{journal}{Journal of Magnetism and Magnetic Materials}}
		\textbf{\bibinfo{volume}{147}}, \bibinfo{pages}{284--298}
		(\bibinfo{year}{1995}).
		\newblock \urlprefix\url{http://dx.doi.org/10.1016/0304-8853(95)00065-8}.
		
		\bibitem{Bryan08}
		\bibinfo{author}{Bryan, M.~T.}, \bibinfo{author}{Schrefl, T.},
		\bibinfo{author}{Atkinson, D.} \& \bibinfo{author}{Allwood, D.~A.}
		\newblock \bibinfo{title}{Magnetic domain wall propagation in nanowires under
			transverse magnetic fields}.
		\newblock \emph{\bibinfo{journal}{Journal of Applied Physics}}
		\textbf{\bibinfo{volume}{103}}, \bibinfo{pages}{073906}
		(\bibinfo{year}{2008}).
		\newblock \urlprefix\url{http://dx.doi.org/10.1063/1.2887918}.
		
		\bibitem{Lu10}
		\bibinfo{author}{Lu, J.} \& \bibinfo{author}{Wang, X.~R.}
		\newblock \bibinfo{title}{Motion of transverse domain walls in thin magnetic
			nanostripes under transverse magnetic fields}.
		\newblock \emph{\bibinfo{journal}{Journal of Applied Physics}}
		\textbf{\bibinfo{volume}{107}}, \bibinfo{pages}{083915}
		(\bibinfo{year}{2010}).
		\newblock \urlprefix\url{http://dx.doi.org/10.1063/1.3386468}.
		
		\bibitem{Goussev13b}
		\bibinfo{author}{Goussev, A.}, \bibinfo{author}{Lund, R.~G.},
		\bibinfo{author}{Robbins, J.~M.}, \bibinfo{author}{Slastikov, V.} \&
		\bibinfo{author}{Sonnenberg, C.}
		\newblock \bibinfo{title}{Fast domain-wall propagation in uniaxial nanowires
			with transverse fields}.
		\newblock \emph{\bibinfo{journal}{Phys. Rev. B}} \textbf{\bibinfo{volume}{88}}
		(\bibinfo{year}{2013}).
		\newblock \urlprefix\url{http://dx.doi.org/10.1103/PhysRevB.88.024425}.
		
		\bibitem{Goehler11}
		\bibinfo{author}{G\"ohler, B.} \emph{et~al.}
		\newblock \bibinfo{title}{Spin selectivity in electron transmission through
			self-assembled monolayers of double-stranded {DNA}}.
		\newblock \emph{\bibinfo{journal}{Science}} \textbf{\bibinfo{volume}{331}},
		\bibinfo{pages}{894--897} (\bibinfo{year}{2011}).
		\newblock \urlprefix\url{http://dx.doi.org/10.1126/science.1199339}.
		
		\bibitem{Naaman12}
		\bibinfo{author}{Naaman, R.} \& \bibinfo{author}{Waldeck, D.~H.}
		\newblock \bibinfo{title}{Chiral-induced spin selectivity effect}.
		\newblock \emph{\bibinfo{journal}{J. Phys. Chem. Lett.}}
		\textbf{\bibinfo{volume}{3}}, \bibinfo{pages}{2178--2187}
		(\bibinfo{year}{2012}).
		\newblock \urlprefix\url{http://dx.doi.org/10.1021/jz300793y}.
		
		\bibitem{Eremko13}
		\bibinfo{author}{Eremko, A.~A.} \& \bibinfo{author}{Loktev, V.~M.}
		\newblock \bibinfo{title}{Spin sensitive electron transmission through helical
			potentials}.
		\newblock \emph{\bibinfo{journal}{Physical Review B}}
		\textbf{\bibinfo{volume}{88}}, \bibinfo{pages}{165409}
		(\bibinfo{year}{2013}).
		\newblock \urlprefix\url{http://dx.doi.org/10.1103/PhysRevB.88.165409}.
		
		\bibitem{Grigoriev07}
		\bibinfo{author}{Grigoriev, S.~V.} \emph{et~al.}
		\newblock \bibinfo{title}{{Principal interactions in the magnetic system
				${\mathrm{Fe}}_{1\ensuremath{-}x}{\mathrm{Co}}_{x}\mathrm{Si}$: Magnetic
				structure and critical temperature by neutron diffraction and SQUID
				measurements}}.
		\newblock \emph{\bibinfo{journal}{Physical Review B}}
		\textbf{\bibinfo{volume}{76}}, \bibinfo{pages}{092407}
		(\bibinfo{year}{2007}).
		\newblock \urlprefix\url{http://dx.doi.org/10.1103/PhysRevB.76.092407}.
		
		\bibitem{Yang15}
		\bibinfo{author}{{Yang}, H.}, \bibinfo{author}{{Thiaville}, A.},
		\bibinfo{author}{{Rohart}, S.}, \bibinfo{author}{{Fert}, A.} \&
		\bibinfo{author}{{Chshiev}, M.}
		\newblock \bibinfo{title}{{Anatomy of Dzyaloshinskii-Moriya Interaction at
				Co/Pt Interfaces}} (\bibinfo{year}{2015}).
		\newblock \eprint{1501.05511}.
		
		\bibitem{Pylypovskyi14}
		\bibinfo{author}{Pylypovskyi, O.~V.}, \bibinfo{author}{Sheka, D.~D.},
		\bibinfo{author}{Kravchuk, V.~P.} \& \bibinfo{author}{Gaididei, Y.}
		\newblock \bibinfo{title}{Effects of surface anisotropy on magnetic vortex
			core}.
		\newblock \emph{\bibinfo{journal}{Journal of Magnetism and Magnetic Materials}}
		\textbf{\bibinfo{volume}{361}}, \bibinfo{pages}{201 -- 205}
		(\bibinfo{year}{2014}).
		\newblock
		\urlprefix\url{http://www.sciencedirect.com/science/article/pii/S0304885314002157}.
		
		\bibitem{btrzx}
		\bibinfo{title}{Bayreuth {U}niversity computing cluster}.
		\newblock \urlprefix\url{http://www.rz.uni-bayreuth.de/}.
		
		\bibitem{unicc}
		\bibinfo{title}{High--performance computing cluster of {T}aras {S}hevchenko
			{N}ational {U}niversity of {K}yiv}.
		\newblock \urlprefix\url{http://cluster.univ.kiev.ua/eng/}.
		
		\bibitem{bitpcluster}
		\bibinfo{title}{Computing grid-cluster of the {B}ogolyubov {I}nsitute for
			{T}heoretical {P}hysics of {NAS} of {U}kraine}.
		\newblock \urlprefix\url{http://horst-7.bitp.kiev.ua}.
		
	\end{thebibliography}
\end{document}